%%%%%%%%%%%%%%%%%%  tex macros for preprints, cm version %%%%%%%%%%%%%%
%         (P. Ginsparg <ginsparg@lanl.gov>, last updated 7/94)
%         hypertex extensions (still provisional), 7/26/94
%	  Some modifications by C.R.Mafra, 2012

%comment out this line to restore non-hyper functionality
%\input hyperbasics

\input amssym.tex % for blackboard bold

\def\unredoffs{}
\tolerance=1000\hfuzz=2pt
\catcode`\@=11 % This allows us to modify PLAIN macros.
\ifx\hyperdef\UNd@FiNeD\def\hyperdef#1#2#3#4{#4}\def\hyperref#1#2#3#4{#4}\def\href#1#2{#2}\fi
\magnification=1200\unredoffs\baselineskip=16pt plus 2pt minus 1pt
\def\Date#1{\vfill\leftline{#1}\tenpoint\supereject%
\footline={\hss\tenrm\hyperdef\hypernoname{page}\folio\folio\hss}}%
% (restores pagenumbers)

%%%%%% Hour:Minute %%%%%%%%%%%%%%%%%
{\count255=\time\divide\count255 by 60 \xdef\hourmin{\number\count255}
 \multiply\count255 by-60\advance\count255 by\time
 \xdef\hourmin{\hourmin:\ifnum\count255<10 0\fi\the\count255}
}
\def\date{\number\day.\number\month.\number\year\ at \hourmin}

%%%%%%%%%%%% Draft mode %%%%%%%%%%%%%
% puts date/time on each page in big mode, writes labels in margins

% use \nolabels to get rid of eqn, ref, and fig labels in draft mode
\def\nolabels{\def\wrlabeL##1{}\def\eqlabeL##1{}\def\reflabeL##1{}}
\def\writelabels{\def\wrlabeL##1{\leavevmode\vadjust{\rlap{\smash%
{\line{{\escapechar=` \hfill\rlap{\sevenrm\hskip.03in\string##1}}}}}}}%
\def\eqlabeL##1{{\escapechar-1\rlap{\sevenrm\hskip.05in\string##1}}}%
\def\reflabeL##1{\noexpand\llap{\noexpand\sevenrm\string\string\string##1}}}
\nolabels

% tagged sec numbers
\global\newcount\secno \global\secno=0
\global\newcount\meqno \global\meqno=1
\def\s@csym{}

%%%%%%%%% Section %%%%%%%%%%%%%
\def\newsec#1\par{\global\advance\secno by1%
{\toks0{#1}\message{(\the\secno. \the\toks0)}}%
\global\subsecno=0\eqnres@t\let\s@csym\secsym\xdef\secn@m{\the\secno}\noindent
{\bf\hyperdef\hypernoname{section}{\the\secno}{\the\secno.} #1}%
\writetoca{{\string\hyperref{}{section}{\the\secno}{\bf \the\secno\quad}} {\bf #1}}\par%
\nobreak\medskip\nobreak\noindent\ignorespaces}
\def\eqnres@t{\xdef\secsym{\the\secno.}\global\meqno=1\bigbreak\bigskip}
\def\sequentialequations{\def\eqnres@t{\bigbreak}}\xdef\secsym{}

%%%%%%%% Subsection %%%%%%%%%%%
\global\newcount\subsecno \global\subsecno=0
\def\subsec#1\par{\global\advance\subsecno by1%
{\toks0{#1}\message{(\s@csym\the\subsecno. \the\toks0)}}%
\global\subsubsecno=0%
\ifnum\lastpenalty>9000\else\bigbreak\fi
\noindent{\it\hyperdef\hypernoname{subsection}{\secn@m.\the\subsecno}%
{\secn@m.\the\subsecno.} #1}\writetoca{\string\hskip1.45cm
{\string\hyperref{}{subsection}{\secn@m.\the\subsecno}{\secn@m.\the\subsecno.}}
{#1}}\par\nobreak\medskip\nobreak\noindent\ignorespaces}

%%%%%%%%%%%%%%% Subsubsection %%%%%%%%%%%%%%%%%%%%%%%%%%%%%%%%%%%%
\global\newcount\subsubsecno \global\subsubsecno=0
\def\subsubsec#1\par{\global\advance\subsubsecno by1%
{\toks0{#1}\message{(\secn@m.\the\subsecno.\the\subsubsecno. \the\toks0)}}%
\global\subsubsubsecno=0%
\ifnum\lastpenalty>9000\else\bigbreak\fi
\noindent{\it\hyperdef\hypernoname{subsubsection}{\secn@m.\the\subsecno\the\subsubsecno}%
{\secn@m.\the\subsecno.\the\subsubsecno.} #1}
%%% Add Subsubsections to Index
%\writetoca{\string\quad{\string\hyperref{}{subsubsection}{\the\secno\the\subsecno\the
%\subsubsecno}{\baselineskip=9pt\it\the\secno.\the\subsecno.\the\subsubsecno.}}
% {\baselineskip=9pt\it\ #1}}
\par\nobreak\medskip\nobreak\noindent\ignorespaces}

%%%%%%%%%%%%%%% Subsubsubsection %%%%%%%%%%%%%%%%%%%%%%%%%%%%%%%%%%%%
\global\newcount\subsubsubsecno \global\subsubsubsecno=0
\def\subsubsubsec#1\par{\global\advance\subsubsubsecno by1%
{\toks0{#1}\message{(\secn@m.\the\subsecno.\the\subsubsecno.\the\subsubsubsecno \the\toks0)}}%
\ifnum\lastpenalty>9000\else\bigbreak\fi
\noindent{\it\hyperdef\hypernoname{subsubsection}{\secn@m.\the\subsecno\the\subsubsecno\the\subsubsubsecno}%
{\secn@m.\the\subsecno.\the\subsubsecno.\the\subsubsubsecno.} #1}%
\par\nobreak\medskip\nobreak\noindent\ignorespaces}

%%%%%%%%% sections with automatic labels %%%%%%%%%%%%%%%%%%%%%%%%%%

%%%%%%%%% section with label %%%%%%%%%%%%%%%%%%%%%%%%%%%%%%%%
\def\newnewsec#1#2\par{\global\advance\secno by1%
{\toks0{#2}\message{(\secn@m. \the\toks0)}}%
\global\subsecno=0\global\subsubsecno=0\eqnres@t\let\s@csym\secsym\xdef\secn@m{\the\secno}\noindent
\ifnum\lastpenalty>9000\else\bigbreak\fi
\noindent{\bf\hyperdef\hypernoname{section}{\secn@m}{\secn@m.} #2}%
\writetoca{{\string\hyperref{}{section}{\the\secno}{\bf \the\secno\quad}} {\bf #2}}
%define section label
\DefWarn#1%
\xdef#1{\noexpand\hyperref{}{section}{\the\secno}%
{\the\secno}}\writedef{#1\leftbracket#1}\wrlabeL{#1=#1}%
\par\nobreak\medskip\nobreak\noindent\ignorespaces}

%%%%%%%%% subsection with label %%%%%%%%%%%%%%%%%%%%%%%%%%%%%%%%
\def\newsubsec#1#2\par{\global\advance\subsecno by1%
{\toks0{#2}\message{(\secn@m.\the\subsecno. \the\toks0)}}%
\global\subsubsecno=0%
\ifnum\lastpenalty>9000\else\bigbreak\fi
\noindent{\it\hyperdef\hypernoname{subsection}{\secn@m.\the\subsecno}%
{\secn@m.\the\subsecno.} #2}
%define section label
\DefWarn#1%
\xdef#1{\noexpand\hyperref{}{subsection}{\secn@m.\the\subsecno}%
{\secn@m.\the\subsecno}}\writedef{#1\leftbracket#1}\wrlabeL{#1=#1}%
\writetoca{\string\hskip1.45cm
{\string\hyperref{}{subsection}{\secn@m.\the\subsecno}{\secn@m.\the\subsecno.}}
{#2}}%
\par\nobreak\medskip\nobreak\noindent\ignorespaces}

%%%%%%%%% subsubsection with label %%%%%%%%%%%%%%%%%%%%%%%%%%%%%%%%
\def\newsubsubsec#1#2\par{\global\advance\subsubsecno by1%
{\toks0{#2}\message{(\secn@m.\the\subsecno.\the\subsubsecno. \the\toks0)}}%
\global\subsubsubsecno=0%
\ifnum\lastpenalty>9000\else\bigbreak\fi
\noindent{\it\hyperdef\hypernoname{subsubsection}{\secn@m.\the\subsecno\the\subsubsecno}%
{\secn@m.\the\subsecno.\the\subsubsecno.} #2}
%define section label
\DefWarn#1%
\xdef#1{\noexpand\hyperref{}{subsubsection}{\secn@m.\the\subsecno.\the\subsubsecno}%
{\secn@m.\the\subsecno.\the\subsubsecno}}\writedef{#1\leftbracket#1}\wrlabeL{#1=#1}%
\par\nobreak\medskip\nobreak\noindent\ignorespaces}

%%%%%%%%% subsubsubsection with label %%%%%%%%%%%%%%%%%%%%%%%%%%%%%%%%
\def\newsubsubsubsec#1#2\par{\global\advance\subsubsubsecno by1%
{\toks0{#2}\message{(\secn@m.\the\subsecno.\the\subsubsecno.\the\subsubsubsecno \the\toks0)}}%
\ifnum\lastpenalty>9000\else\bigbreak\fi
\noindent{\it\hyperdef\hypernoname{subsubsection}{\secn@m.\the\subsecno\the\subsubsecno\the\subsubsubsecno}%
{\secn@m.\the\subsecno.\the\subsubsecno.\the\subsubsubsecno.} #2}
%define section label
\DefWarn#1%
\xdef#1{\noexpand\hyperref{}{subsubsubsection}{\secn@m.\the\subsecno.\the\subsubsecno.\the\subsubsubsecno}%
{\secn@m.\the\subsecno.\the\subsubsecno.\the\subsubsubsecno}}\writedef{#1\leftbracket#1}\wrlabeL{#1=#1}%
\par\nobreak\medskip\nobreak\noindent\ignorespaces}

%%%%%%% Appendix %%%%%%%%%%%%%%
\def\appendix#1#2{\global\meqno=1\global\subsecno=0\global\subsubsecno=0\xdef\secsym{\hbox{#1.}}%
\bigbreak\bigskip\noindent{\bf Appendix \hyperdef\hypernoname{appendix}{#1}%
{#1.} #2}{\toks0{(#1. #2)}\message{\the\toks0}}%
\xdef\s@csym{#1.}\xdef\secn@m{#1}%
\writetoca{{\string\hyperref{}{appendix}{#1}{\bf {#1}\quad}} {\bf #2}}%
\par\nobreak\medskip\nobreak}

% \eqn\label{a+b=c}   gives displayed equation, numbered consecutively within sections.
% \eqnn, \eqna        define labels in advance, use \eqna\label before an eqalign and
%                     later \label a, \label b etc inside eqalign to get (2.3a), (2.3b) etc
%
\def\checkm@de#1#2{\ifmmode{\def\f@rst##1{##1}\hyperdef\hypernoname{equation}%
{#1}{#2}}\else\hyperref{}{equation}{#1}{#2}\fi}
\def\eqnn#1{\DefWarn#1\xdef #1{(\noexpand\relax\noexpand\checkm@de%
{\s@csym\the\meqno}{\secsym\the\meqno})}%
\wrlabeL#1\writedef{#1\leftbracket#1}\global\advance\meqno by1}
\def\f@rst#1{\c@t#1a\em@ark}\def\c@t#1#2\em@ark{#1}
\def\eqna#1{\DefWarn#1\wrlabeL{#1$\{\}$}%
\xdef #1##1{(\noexpand\relax\noexpand\checkm@de%
{\s@csym\the\meqno\noexpand\f@rst{##1}1}{\hbox{$\secsym\the\meqno##1$}})}
\writedef{#1\numbersign1\leftbracket#1{\numbersign1}}\global\advance\meqno by1}
\def\eqn#1#2{\DefWarn#1%
\xdef #1{(\noexpand\hyperref{}{equation}{\s@csym\the\meqno}%
{\secsym\the\meqno})}$$#2\eqno(\hyperdef\hypernoname{equation}%
{\s@csym\the\meqno}{\secsym\the\meqno})\eqlabeL#1$$%
\writedef{#1\leftbracket#1}\global\advance\meqno by1}
\def\xeqn{\expandafter\xe@n}\def\xe@n(#1){#1}
\def\xeqna#1{\expandafter\xe@n#1}
\def\eqns#1{(\e@ns #1{\hbox{}})}
\def\e@ns#1{\ifx\UNd@FiNeD#1\message{eqnlabel \string#1 is undefined.}%
\xdef#1{(?.?)}\fi{\let\hyperref=\relax\xdef\next{#1}}%
\ifx\next\em@rk\def\next{}\else%
\ifx\next#1\xeqn#1\else\def\n@xt{#1}\ifx\n@xt\next#1\else\xeqna#1\fi
\fi\let\next=\e@ns\fi\next}
\def\DefWarn#1{}
%
% footnotes
\newskip\footskip\footskip14pt plus 1pt minus 1pt %sets footnote baselineskip
\def\footnotefont{\ninepoint}\def\f@t#1{\footnotefont #1\@foot}
\def\f@@t{\baselineskip\footskip\bgroup\footnotefont\aftergroup\@foot\let\next}
\setbox\strutbox=\hbox{\vrule height9.5pt depth4.5pt width0pt}
\global\newcount\ftno \global\ftno=0
\def\foot{\global\advance\ftno by1\def\foot@rg{\hyperref{}{footnote}%
{\the\ftno}{\the\ftno}\xdef\foot@rg{\noexpand\hyperdef\noexpand\hypernoname%
{footnote}{\the\ftno}{\the\ftno}}}\footnote{$^{\foot@rg}$}}
%
%
%     \ref\label{text}
% generates a number, assigns it to \label, generates an entry.
% To list the refs on a separate page,  \listrefs
%
\global\newcount\refno \global\refno=1
\newwrite\rfile
\def\ref{[\hyperref{}{reference}{\the\refno}{\the\refno}]\nref}
\def\nref#1{\DefWarn#1%
\xdef#1{[\noexpand\hyperref{}{reference}{\the\refno}{\the\refno}]}%
\writedef{#1\leftbracket#1}%
\ifnum\refno=1\immediate\openout\rfile=\jobname.refs\fi
\chardef\wfile=\rfile\immediate\write\rfile{\noexpand\item{[\noexpand\hyperdef%
\noexpand\hypernoname{reference}{\the\refno}{\the\refno}]\ }%
\reflabeL{#1\hskip.31in}\pctsign}\global\advance\refno by1\findarg}
%	horrible hack to sidestep tex \write limitation
\def\findarg#1#{\begingroup\obeylines\newlinechar=`\^^M\pass@rg}
{\obeylines\gdef\pass@rg#1{\writ@line\relax #1^^M\hbox{}^^M}%
\gdef\writ@line#1^^M{\expandafter\toks0\expandafter{\striprel@x #1}%
\edef\next{\the\toks0}\ifx\next\em@rk\let\next=\endgroup\else\ifx\next\empty%
\else\immediate\write\wfile{\the\toks0}\fi\let\next=\writ@line\fi\next\relax}}
\def\striprel@x#1{} \def\em@rk{\hbox{}}
\def\lref{\begingroup\obeylines\lr@f}
\def\lr@f#1#2{\DefWarn#1\gdef#1{\let#1=\UNd@FiNeD\ref#1{#2}}\endgroup\unskip}

\def\addref#1{\immediate\write\rfile{\noexpand\item{}#1}} %now unnecessary
\def\listrefs{\vfill\supereject\immediate\closeout\rfile\writestoppt
\baselineskip=\footskip\centerline{{\bf References}}\bigskip{\parindent=20pt%
\frenchspacing\escapechar=` \input \jobname.refs\vfill\eject}\nonfrenchspacing}
\def\startrefs#1{\immediate\openout\rfile=\jobname.refs\refno=#1}
\def\xref{\expandafter\xr@f}\def\xr@f[#1]{#1}
\def\refs#1{\count255=1[\r@fs #1{\hbox{}}]}
\def\r@fs#1{\ifx\UNd@FiNeD#1\message{reflabel \string#1 is undefined.}%
\nref#1{need to supply reference \string#1.}\fi%
\vphantom{\hphantom{#1}}{\let\hyperref=\relax\xdef\next{#1}}%
\ifx\next\em@rk\def\next{}%
\else\ifx\next#1\ifodd\count255\relax\xref#1\count255=0\fi%
\else#1\count255=1\fi\let\next=\r@fs\fi\next}
%

%
% this is ugly, but moore insists
\newwrite\ffile\global\newcount\figno \global\figno=1
\def\fig{fig.~\hyperref{}{figure}{\the\figno}{\the\figno}\nfig}
\def\nfig#1{\DefWarn#1%
\xdef#1{fig.~\noexpand\hyperref{}{figure}{\the\figno}{\the\figno}}%
\writedef{#1\leftbracket fig.\noexpand~\xfig#1}%
\ifnum\figno=1\immediate\openout\ffile=\jobname.figs\fi\chardef\wfile=\ffile%
{\let\hyperref=\relax
\immediate\write\ffile{\noexpand\medskip\noexpand\item{Fig.\ %
\noexpand\hyperdef\noexpand\hypernoname{figure}{\the\figno}{\the\figno}. }
\reflabeL{#1\hskip.55in}\pctsign}}\global\advance\figno by1\findarg}
\def\xfig{\expandafter\xf@g}\def\xf@g fig.\penalty\@M\ {}
\def\figs#1{figs.~\f@gs #1{\hbox{}}}
\def\f@gs#1{{\let\hyperref=\relax\xdef\next{#1}}\ifx\next\em@rk\def\next{}\else
\ifx\next#1\xfig #1\else#1\fi\let\next=\f@gs\fi\next}
%
%% because TeXlive 2011 is buggy wrt to tikz pictures with plain TeX..
\def\figin{\epsfcheck\figin}\def\figins{\epsfcheck\figins}
\def\epsfcheck{\ifx\epsfbox\UnDeFiNeD
\message{(NO epsf.tex, FIGURES WILL BE IGNORED)}
\gdef\figin##1{\vskip2in}\gdef\figins##1{\hskip.5in}% blank space instead
\else\message{(FIGURES WILL BE INCLUDED)}%
\gdef\figin##1{##1}\gdef\figins##1{##1}\fi}
\def\figinsert{\goodbreak\topinsert}
\def\ifig#1#2#3{\DefWarn#1\xdef#1{fig.~\the\figno}
\writedef{#1\leftbracket fig.\noexpand~\the\figno}%
\figinsert\figin{\centerline{#3}}
\smallskip
\leftskip=0pt \rightskip=0pt
\baselineskip12pt\noindent
{{\bf Fig.~\the\figno}\ \ninepoint #2}
\medskip
\global\advance\figno by1\par\endinsert}
%%%%%%%%%%%%%%%%%%%%%%%%%%%%%%%%%%%%%%%%%%%%%%%%%%%%%%%%%
\newwrite\lfile
{\escapechar-1\xdef\pctsign{\string\%}\xdef\leftbracket{\string\{}
\xdef\rightbracket{\string\}}\xdef\numbersign{\string\#}}
\def\writedefs{\immediate\openout\lfile=label.defs \def\writedef##1{%
{\let\hyperref=\relax\let\hyperdef=\relax\let\hypernoname=\relax
 \immediate\write\lfile{\string\checkdef\string##1\rightbracket}}}}%
\def\writestop{\def\writestoppt{\immediate\write\lfile{\string\pageno
 \the\pageno\string\startrefs\leftbracket\the\refno\rightbracket
 \string\def\string\secsym\leftbracket\secsym\rightbracket
 \string\secno\the\secno\string\meqno\the\meqno}\immediate\closeout\lfile}}
\def\writestoppt{}\def\writedef#1{}

% Section, subsection and appendix labels %
% Note that there must be a blanck line after \newsec,\subsec and before \seclab,\subseclab!
\def\seclab#1\par{\DefWarn#1%
\xdef #1{\noexpand\hyperref{}{section}{\the\secno}{\the\secno}}%
\writedef{#1\leftbracket#1}\wrlabeL{#1=#1}\par%
\nobreak\medskip\nobreak\noindent\ignorespaces}
\def\subseclab#1\par{\DefWarn#1%
\xdef #1{\noexpand\hyperref{}{subsection}{\the\secno.\the\subsecno}%
{\the\secno.\the\subsecno}}\writedef{#1\leftbracket#1}\wrlabeL{#1=#1}\par%
\nobreak\medskip\nobreak\noindent\ignorespaces}
\def\subsubseclab#1\par{\DefWarn#1%
\xdef#1{\noexpand\hyperref{}{subsubsection}{\the\secno.\the\subsecno.\the\subsubsecno}%
{\the\secno.\the\subsecno.\the\subsubsecno}}\writedef{#1\leftbracket#1}\wrlabeL{#1=#1}\par%
\nobreak\medskip\nobreak\noindent\ignorespaces}
\def\applab#1\par{\DefWarn#1%
\xdef#1{\noexpand\hyperref{}{appendix}{\secn@m}{\secn@m}}%
\writedef{#1\leftbracket#1}\wrlabeL{#1=#1}%
\par\nobreak\medskip\nobreak\noindent\ignorespaces}
\def\appsublab#1{\DefWarn#1%
\xdef #1{\noexpand\hyperref{}{appendix}{\secn@m.\the\subsecno}{\secn@m.\the\subsecno}}%
\writedef{#1\leftbracket#1}\wrlabeL{#1=#1}}
\newwrite\tfile \def\writetoca#1{}
\def\leaderfill{\leaders\hbox to 1em{\hss.\hss}\hfill}
% use this to write file with table of contents
\def\writetoc{\immediate\openout\tfile=\jobname.toc
   \def\writetoca##1{{\edef\next{\write\tfile{\noindent ##1
   \string\leaderfill{
% comment this line if you don't want hyperlinked page numbers on TOC
   \string\hyperref{}{page}{\noexpand\number\pageno}%
   {\noexpand\number\pageno}} \par}}\next}}
}
% and this lists table of contents on second pass
\newread\ch@ckfile
\def\listtoc{\immediate\closeout\tfile\immediate\openin\ch@ckfile=\jobname.toc
\ifeof\ch@ckfile\message{no file \jobname.toc, no table of contents this pass}%
\else\closein\ch@ckfile\centerline{\bf Contents}\nobreak\medskip%
{\baselineskip=15.5pt\footnotefont\parskip=0pt\catcode`\@=11\input\jobname.toc
\catcode`\@=12\bigbreak\bigskip}\fi}
\catcode`\@=12 % at signs are no longer letters
\def\tenpoint{\def\rm{\fam0\tenrm}% switch back to 10-point type
\textfont0=\tenrm \scriptfont0=\sevenrm \scriptscriptfont0=\fiverm
\textfont1=\teni  \scriptfont1=\seveni  \scriptscriptfont1=\fivei
\textfont2=\tensy \scriptfont2=\sevensy \scriptscriptfont2=\fivesy
\textfont\itfam=\tenit \def\it{\fam\itfam\tenit}\def\footnotefont{\ninepoint}%
\textfont\bffam=\tenbf \def\bf{\fam\bffam\tenbf}\def\sl{\fam\slfam\tensl}\rm}
\font\ninerm=cmr9 \font\sixrm=cmr6 \font\ninei=cmmi9 \font\sixi=cmmi6
\font\ninesy=cmsy9 \font\sixsy=cmsy6 \font\ninebf=cmbx9
\font\nineit=cmti9 \font\ninesl=cmsl9 \skewchar\ninei='177
\skewchar\sixi='177 \skewchar\ninesy='60 \skewchar\sixsy='60
\def\ninepoint{\def\rm{\fam0\ninerm}% switch to footnote font
\textfont0=\ninerm \scriptfont0=\sixrm \scriptscriptfont0=\fiverm
\textfont1=\ninei \scriptfont1=\sixi \scriptscriptfont1=\fivei
\textfont2=\ninesy \scriptfont2=\sixsy \scriptscriptfont2=\fivesy
\textfont\itfam=\ninei \def\it{\fam\itfam\nineit}\def\sl{\fam\slfam\ninesl}%
\textfont\bffam=\ninebf \def\bf{\fam\bffam\ninebf}\rm}
%
%---------------------------------------------------------------------
\hyphenation{anom-aly anom-alies coun-ter-term coun-ter-terms}

% Caption for inline tikzpictures
%\def\DefWarn#1{}
\def\tikzcaption#1#2{\DefWarn#1\xdef#1{Fig.~\the\figno}
\writedef{#1\leftbracket Fig.\noexpand~\the\figno}%
{
\smallskip
\leftskip=20pt \rightskip=20pt \baselineskip12pt\noindent
{{\bf Fig.~\the\figno}\ \ninepoint #2}
\bigskip
\global\advance\figno by1 \par}}

% convert numbers [1-9] to upper case letters [A-I]
\def\ntoalpha#1{%
\ifcase#1%
@%
\or A\or B\or C\or D\or E\or F\or G\or H\or I\or J\or K\or L\or M%
\fi
}

% Appendix label
\global\newcount\appno \global\appno=1
\def\applab#1{\xdef #1{\ntoalpha{\appno}}\writedef{#1\leftbracket#1}\wrlabeL{#1=#1}
\global\advance\appno by1}

% Clean up the title page definitions
\def\preprint#1 #2\par{\rightline{\vbox{\baselineskip12pt\hbox{#1}\hbox{#2}}}\vskip2cm}
% title with more than one line (note the blanck line in between)
%\title some line
%
%\tile another line
\def\title#1\par{\centerline{\bf #1}\nopagenumbers\pageno=0}
\def\author#1\par{\bigskip\bigskip\centerline{#1}}

\newcount\addressno

\def\email#1#2{%\unskip$^#1$
\footnote{\null}{\kern-\parindent \llap{$^#1$\hskip1pt}email: #2}}

% centermode for address lines
\def\startcenter{%
  \par
  \begingroup
  \leftskip=0pt plus 1fil
  \rightskip=\leftskip
  \parindent=0pt
  \parfillskip=0pt
}
\def\stopcenter{\endgroup}

\def\address{\bigskip%
  \ifnum\the\addressno=0\else\stopcenter\endgroup\fi
  \advance\addressno by 1%
  \begingroup
  \startcenter
  \it
  \obeylines
  \addressAux
}
\def\addressAux#1{#1}

% need to stop center mode and obeylines from address
\def\abstract{\stopcenter\endgroup\bigskip\bigskip\noindent}

% some sample definitions
\def\Dsl{\,\raise.15ex\hbox{/}\mkern-13.5mu D} %this one can be subscripted
\def\dsl{\raise.15ex\hbox{/}\kern-.57em\partial}
 \def\Tr{{\rm Tr}}
\def\boxeqn#1{\vcenter{\vbox{\hrule\hbox{\vrule\kern3pt\vbox{\kern3pt
	\hbox{${\displaystyle #1}$}\kern3pt}\kern3pt\vrule}\hrule}}}

 %pound sterling

\def\half{{1\over 2}}

\def\bar{\overline}
\def\({\left(}
\def\){\right)}

% blackboard bold
\def\bV{{\Bbb V}}
\def\bA{{\Bbb A}}
\def\bW{{\Bbb W}}
\def\bF{{\Bbb F}}

\def\Box{\square}

% primed summation symbol

% length of words, |P|

 %redefine plain TeX \Im..
% small inlined fractions, from the TeXbook
\def\sfrac#1/#2{\kern.1em\raise.5ex\hbox{\the\scriptfont0 #1}%
\kern-.1em/\kern-.15em\lower.25ex\hbox{\the\scriptfont0 #2}}

%shuffle product

%\owedge

% From Knuth's \pfbox macro
\def\qed{\hbox{\hskip 3pt
%\lower2pt
\vbox{\hrule\hbox to 7pt{\vrule height 7pt\hfill\vrule}
\hrule}}\hskip3pt}

% do not display overfull marks
\overfullrule=0pt\relax

\frenchspacing

% DefWarn-like behavior for labels defined in advance
\def\checkdef#1#2{
\ifx\UndeFined#1%
	\def#1{#2}
%\immediate\write16{*** define label \string#1 by #2 ***}
\else
	\immediate\write16{*** BUG ***: the label \string#1 is already defined ***}
\fi
}
% define labels in advance
\newread\instream

\openin\instream= label.defs
\ifeof\instream\message{No labels in advance yet. Wait till next pass.}
\else\closein\instream \input label.defs
\fi
\writedefs

%%% References with hyperlinks to arxiv.org; both styles accepted
% Change arXiv to \arXiv ie
% [arXiv:hep-th/1234567].     --> [\arXiv:hep-th/1234567].
% [arXiv:1234.5678 [hep-th]]. --> [\arXiv:1234.5678 [hep-th]].
% Need to strip trailing [hep-th] (if present) to define valid URL
\def\arXiv:#1].{\hepthStrip#1 \nil}
\def\hepthStrip#1 #2\nil{\href{http://arxiv.org/abs/#1}{arXiv:#1 #2\unskip}].}

%%% Fraktur fonts for Berends-Giele components

% Caption for inline tikzpictures
%\def\DefWarn#1{}
\def\tikzcaption#1#2{\DefWarn#1\xdef#1{Fig.~\the\figno}
\writedef{#1\leftbracket Fig.\noexpand~\the\figno}%
{
\smallskip
\leftskip=20pt \rightskip=20pt \baselineskip12pt\noindent
{{\bf Fig.~\the\figno}\ \ninepoint #2}
\bigskip
\global\advance\figno by1 \par}}

% convert numbers [1-9] to upper case letters [A-I]
\def\ntoalpha#1{%
\ifcase#1%
@%
\or A\or B\or C\or D\or E\or F\or G\or H\or I\or J\or K\or L\or M%
\fi
}

% Appendix label
\global\newcount\appno \global\appno=1
\def\applab#1{\xdef #1{\ntoalpha{\appno}}\writedef{#1\leftbracket#1}\wrlabeL{#1=#1}
\global\advance\appno by1}

% Clean up the title page definitions
\def\preprint#1 #2\par{\rightline{\vbox{\baselineskip12pt\hbox{#1}\hbox{#2}}}\vskip2cm}
% title with more than one line (note the blanck line in between)
%\title some line
%
%\tile another line
\def\title#1\par{\centerline{\bf #1}\nopagenumbers\pageno=0}
\def\author#1\par{\bigskip\bigskip\centerline{#1}}

\newcount\addressno

\def\email#1#2{%\unskip$^#1$
\footnote{\null}{\kern-\parindent \llap{$^#1$\hskip1pt}email: #2}}

% centermode for address lines
\def\startcenter{%
  \par
  \begingroup
  \leftskip=0pt plus 1fil
  \rightskip=\leftskip
  \parindent=0pt
  \parfillskip=0pt
}
\def\stopcenter{\endgroup}

\def\address{\bigskip%
  \ifnum\the\addressno=0\else\stopcenter\endgroup\fi
  \advance\addressno by 1%
  \begingroup
  \startcenter
  \it
  \obeylines
  \addressAux
}
\def\addressAux#1{#1}

% need to stop center mode and obeylines from address
\def\abstract{\stopcenter\endgroup\bigskip\bigskip\noindent}

% some sample definitions
\def\Dsl{\,\raise.15ex\hbox{/}\mkern-13.5mu D} %this one can be subscripted
\def\dsl{\raise.15ex\hbox{/}\kern-.57em\partial}
 \def\Tr{{\rm Tr}}
\def\boxeqn#1{\vcenter{\vbox{\hrule\hbox{\vrule\kern3pt\vbox{\kern3pt
	\hbox{${\displaystyle #1}$}\kern3pt}\kern3pt\vrule}\hrule}}}

 %pound sterling

\def\half{{1\over 2}}

\def\bar{\overline}
\def\({\left(}
\def\){\right)}

% blackboard bold
\def\bV{{\Bbb V}}
\def\bU{{\Bbb U}}
\def\bA{{\Bbb A}}

\def\bW{{\Bbb W}}
\def\bF{{\Bbb F}}

\def\bZ{{\Bbb Z}}

\def\Box{\square}

% primed summation symbol

% length of words, |P|

 %redefine plain TeX \Im..
% small inlined fractions, from the TeXbook
\def\sfrac#1/#2{\kern.1em\raise.5ex\hbox{\the\scriptfont0 #1}%
\kern-.1em/\kern-.15em\lower.25ex\hbox{\the\scriptfont0 #2}}

%shuffle product

%\owedge

% From Knuth's \pfbox macro
\def\qed{\hbox{\hskip 3pt
%\lower2pt
\vbox{\hrule\hbox to 7pt{\vrule height 7pt\hfill\vrule}
\hrule}}\hskip3pt}

% do not display overfull marks
\overfullrule=0pt\relax

\frenchspacing

% DefWarn-like behavior for labels defined in advance
\def\checkdef#1#2{
\ifx\UndeFined#1%
	\def#1{#2}
%\immediate\write16{*** define label \string#1 by #2 ***}
\else
	\immediate\write16{*** BUG ***: the label \string#1 is already defined ***}
\fi
}
\input epsf.tex

\def\frac#1#2{{#1\over #2}}

 % local nabla

\def\centretable#1{ \hbox to \hsize {\hfill\vbox{
                    \offinterlineskip \tabskip=0pt \halign{#1} }\hfill} }

\preprint{}

\vskip -.5in

\title The pure spinor superparticle and 10D super-Yang-Mills amplitudes

\author
Max Guillen\email{{\dagger\ddagger*}}{maxgui@chalmers.se}$^{\dagger\ddagger*}$,
Marcelo dos Santos\email{{\#*}}{mafsantos@ucdavis.edu}$^{\#*}$ and
Eggon Viana\email{{\star\circ*}}{eggon.viana@unesp.br}$^{\star\circ*}$

\address
${\dagger}$ Department of Physics and Astronomy, Uppsala University, 75108 Uppsala, Sweden

\vskip .1in
$\ddagger$ Department of Mathematical Sciences, Chalmers University of Technology and the University of Gothenburg, SE-412 96 Gothenburg, Sweden

\vskip .1in
${*}$ ICTP South American Institute for Fundamental Research
Instituto de F\'{i}sica Te\'{o}rica, UNESP-Universidade Estadual Paulista
R. Dr. Bento T. Ferraz 271, Bl. II, S\~{a}o Paulo 01140-070, SP, Brazil

\vskip .1in
${\#}$ Center for Quantum Mathematics and Physics (QMAP)
Department of Physics \& Astronomy, University of California, Davis, CA 95616 USA

\vskip .1in
${\star}$ Instituto Gallego de F\'{i}sica de Altas Energ\'{i}as (IGFAE), Spain

\vskip .1in
${\circ}$ Department of Mathematical Sciences, Durham University, Durham DH1 3LE, UK

\vskip -.1in

\abstract

We present a prescription for computing tree-level scattering amplitudes in 10D super-Yang-Mills (SYM) 
theory using the pure spinor worldline formalism. The pure spinor formalism has proven to be a 
powerful framework for studying supersymmetric field theories, providing manifestly covariant and 
BRST-invariant formulations of amplitudes. By incorporating the worldline approach, we construct a 
first-quantized representation of SYM amplitudes in 10D, where interactions are encoded through the 
insertion of vertex operators along the particle’s trajectory. 
We explicitly compute the N-point function, demonstrating an agreement with the $\alpha'\to0$ 
limit of open superstring amplitudes and confirming that the kinematic numerators satisfy the 
expected BRST relations. Our results establish the pure spinor worldline formalism as a tool 
for studying scattering amplitudes and suggest further applications to 11D supergravity.

\bigskip
\bigskip
\bigskip
\bigskip
\vskip -.2in
\Date {August 2025}

%**************************************

\lref\ICTP{
N.~Berkovits,
``ICTP lectures on covariant quantization of the superstring,''
ICTP Lect. Notes Ser. 13 (2003) 57 [hep-th/0209059].
}
\lref\pselevenparticle{
%\cite{Guillen:2017mte}
M.~Guillen,
``Equivalence of the 11D pure spinor and Brink-Schwarz-like superparticle cohomologies,''
Phys. Rev. D {\bf 97}, no.6, 066002 (2018).
[arXiv:1705.06316 [hep-th]].
%5 citations counted in INSPIRE as of 13 Dec 2022
}

\lref\BerkovitsPS{
N.~Berkovits,
``Super Poincare covariant quantization of the superstring,''
JHEP {\bf 04}, 018 (2000).
[arXiv:hep-th/0001035 [hep-th]].
%576 citations counted in INSPIRE as of 20 Jul 2021
}

\lref\pssupermembrane{
N.~Berkovits,
``Towards covariant quantization of the supermembrane,''
JHEP {\bf 09}, 051 (2002).
[arXiv:hep-th/0201151 [hep-th]].
%99 citations counted in INSPIRE as of 13 Dec 2022
}

\lref\neight{
M.~Cederwall,
``N=8 superfield formulation of the Bagger-Lambert-Gustavsson model,''
JHEP {\bf 09}, 116 (2008).
[arXiv:0808.3242 [hep-th]].
%39 citations counted in INSPIRE as of 13 Dec 2022
}

\lref\nsix{
M.~Cederwall,
``Superfield actions for N=8 and N=6 conformal theories in three dimensions,''
JHEP {\bf 10}, 070 (2008).
[arXiv:0809.0318 [hep-th]].
%37 citations counted in INSPIRE as of 13 Dec 2022
}

\lref\nfour{
M.~Cederwall,
``An off-shell superspace reformulation of D=4, N=4 super-Yang-Mills theory,''
Fortsch. Phys. {\bf 66}, no.1, 1700082 (2018).
[arXiv:1707.00554 [hep-th]].
%6 citations counted in INSPIRE as of 13 Dec 2022
}

\lref\pssugra{
M.~Cederwall,
``D=11 supergravity with manifest supersymmetry,''
Mod. Phys. Lett. A {\bf 25}, 3201-3212 (2010).
[arXiv:1001.0112 [hep-th]].
%50 citations counted in INSPIRE as of 13 Dec 2022
}

\lref\psborninfeld{
M.~Cederwall and A.~Karlsson,
``Pure spinor superfields and Born-Infeld theory,''
JHEP {\bf 11}, 134 (2011).
[arXiv:1109.0809 [hep-th]].
%26 citations counted in INSPIRE as of 13 Dec 2022
}

\lref\mafraone{
N.~Berkovits and C.~R.~Mafra,
``Some Superstring Amplitude Computations with the Non-Minimal Pure Spinor Formalism,''
JHEP {\bf 11}, 079 (2006).
[arXiv:hep-th/0607187 [hep-th]].
%77 citations counted in INSPIRE as of 13 Dec 2022
}

\lref\mafratwo{
H.~Gomez and C.~R.~Mafra,
``The Overall Coefficient of the Two-loop Superstring Amplitude Using Pure Spinors,''
JHEP {\bf 05}, 017 (2010).
[arXiv:1003.0678 [hep-th]].
%44 citations counted in INSPIRE as of 13 Dec 2022
}

\lref\mafrathree{
H.~Gomez and C.~R.~Mafra,
``The closed-string 3-loop amplitude and S-duality,''
JHEP {\bf 10}, 217 (2013).
[arXiv:1308.6567 [hep-th]].
%95 citations counted in INSPIRE as of 13 Dec 2022
}

\lref\rnspsone{
N.~Berkovits,
``Covariant Map Between Ramond-Neveu-Schwarz and Pure Spinor Formalisms for the Superstring,''
JHEP {\bf 04}, 024 (2014).
[arXiv:1312.0845 [hep-th]].
%6 citations counted in INSPIRE as of 13 Dec 2022
}

\lref\rnspstwo{
N.~Berkovits,
``Manifest spacetime supersymmetry and the superstring,''
JHEP {\bf 10}, 162 (2021).
[arXiv:2106.04448 [hep-th]].
%6 citations counted in INSPIRE as of 13 Dec 2022
}

\lref\maxmaor{
M.~Ben-Shahar and M.~Guillen,
``10D super-Yang-Mills scattering amplitudes from its pure spinor action,''
JHEP {\bf 12}, 014 (2021).
[arXiv:2108.11708 [hep-th]].
%9 citations counted in INSPIRE as of 13 Dec 2022
}

\lref\maxmaorelevend{
M.~Ben-Shahar and M.~Guillen,
``Superspace expansion of the 11D linearized superfields in the pure spinor formalism, and the covariant vertex operator,''
JHEP 09, 018 (2023). 
[arXiv:2305.19898 [hep-th]].
}

\lref\MafraNpoint{
C.~R.~Mafra, O.~Schlotterer, and S.~Stieberger,
``Complete N-Point Superstring Disk Amplitude I. Pure Spinor Computation''
Nucl. Phys. B {\bf 873}, 419-460 (2013).
[arXiv:1106.2645 [hep-th]].
%194 citations counted in INSPIRE as of 13 Dec 2022
}

\lref\mafraoli{
C.~R.~Mafra and O.~Schlotterer,
``Multiparticle SYM equations of motion and pure spinor BRST blocks,''
JHEP {\bf 07}, 153 (2014).
[arXiv:1404.4986 [hep-th]].
%70 citations counted in INSPIRE as of 13 Dec 2022
}

\lref\dynamical{
N.~Berkovits,
``Dynamical twisting and the b ghost in the pure spinor formalism,''
JHEP {\bf 06}, 091 (2013).
[arXiv:1305.0693 [hep-th]].
}

\lref\xiyin{
C.~M.~Chang, Y.~H.~Lin, Y.~Wang, and X.~Yin,
``Deformations with Maximal Supersymmetries Part 2: Off-shell Formulation,''
JHEP {\bf 04}, 171 (2016).
[arXiv:1403.0709 [hep-th]].
%15 citations counted in INSPIRE as of 13 Dec 2022
}

\lref\chiralmax{
M.~Guillen,
``Green-Schwarz and pure spinor formulations of chiral strings,''
JHEP {\bf 12}, 029 (2021).
[arXiv:2108.11724 [hep-th]].
%1 citations counted in INSPIRE as of 13 Dec 2022
}

\lref\bcjone{
Z.~Bern, J.~J.~M.~Carrasco, and H.~Johansson,
``New Relations for Gauge-Theory Amplitudes,''
Phys. Rev. D {\bf 78}, 085011 (2008).
[arXiv:0805.3993 [hep-ph]].
%1030 citations counted in INSPIRE as of 13 Dec 2022
}

\lref\bcjtwo{
Z.~Bern, J.~J.~M.~Carrasco, and H.~Johansson,
``Perturbative Quantum Gravity as a Double Copy of Gauge Theory,''
Phys. Rev. Lett. {\bf 105}, 061602 (2010).
[arXiv:1004.0476 [hep-th]].
%743 citations counted in INSPIRE as of 13 Dec 2022
}

\lref\bcjthree{
Z.~Bern, J.~J.~Carrasco, M.~Chiodaroli, H.~Johansson, and R.~Roiban,
``The Duality Between Color and Kinematics and its Applications,''
[arXiv:1909.01358 [hep-th]].
}

\lref\maximalloopcederwall{
M.~Cederwall and A.~Karlsson,
``Loop amplitudes in maximal supergravity with manifest supersymmetry,''
JHEP {\bf 03}, 114 (2013).
[arXiv:1212.5175 [hep-th]].
%20 citations counted in INSPIRE as of 13 Dec 2022
}

\lref\maxnotesworldline{
M.~Guillen,
``Notes on the 11D pure spinor wordline vertex operators,''
JHEP {\bf 08}, 122 (2020).
[arXiv:2006.06022 [hep-th]].
%3 citations counted in INSPIRE as of 13 Dec 2022
}

\lref\OdaTonin{
I.~Oda and M.~Tonin,
``On the Berkovits covariant quantization of GS superstring,''
Phys. Lett. B {\bf 520}, 398-404 (2001).
[arXiv:hep-th/0109051 [hep-th]].
%65 citations counted in INSPIRE as of 20 Jul 2021
}

\lref\perturbiner{
A.~A.~Rosly and K.~G.~Selivanov,
``On amplitudes in selfdual sector of Yang-Mills theory,''
Phys. Lett. B {\bf 399}, 135-140 (1997).
[arXiv:hep-th/9611101 [hep-th]].
%68 citations counted in INSPIRE as of 13 Dec 2022
}
\lref\NMPS{
	N.~Berkovits,
	``Pure spinor formalism as an N=2 topological string,''
	JHEP {\bf 0510}, 089 (2005).
	[hep-th/0509120].
	%%CITATION = hep-th/0509120%%
}

\lref\elevendsimplifiedb{
N.~Berkovits and M.~Guillen,
``Simplified $D = 11$ pure spinor $b$ ghost,''
JHEP {\bf 07}, 115 (2017).
[arXiv:1703.05116 [hep-th]].
%3 citations counted in INSPIRE as of 13 Dec 2022
}
\lref\brinkschwarz{
L.~Brink and J.~H.~Schwarz,
``Quantum Superspace,''
Phys. Lett. B {\bf 100}, 310-312 (1981).
%435 citations counted in INSPIRE as of 13 Dec 2022
}

\lref\brinkhowe{
L.~Brink and P.~S.~Howe,
``Eleven-Dimensional Supergravity on the Mass-Shell in Superspace,''
Phys. Lett. B {\bf 91}, 384-386 (1980).
%170 citations counted in INSPIRE as of 13 Dec 2022
}

\lref\quartet{
T.~Kugo and I.~Ojima,
``Local Covariant Operator Formalism of Nonabelian Gauge Theories and Quark Confinement Problem,''
Prog. Theor. Phys. Suppl. {\bf 66}, 1-130 (1979).
}

\lref\cederwallequations{
N.~Berkovits and M.~Guillen,
``Equations of motion from Cederwall's pure spinor superspace actions,''
JHEP {\bf 08}, 033 (2018).
[arXiv:1804.06979 [hep-th]].
}

\lref\pssreview{
M.~Cederwall,
``Pure spinor superfields -- an overview,''
Springer Proc. Phys. {\bf 153}, 61-93 (2014).
[arXiv:1307.1762 [hep-th]].
%35 citations counted in INSPIRE as of 15 Dec 2022
}

\lref\tendsupertwistors{
N.~Berkovits,
``Ten-Dimensional Super-Twistors and Super-Yang-Mills,''
JHEP {\bf 04}, 067 (2010).
[arXiv:0910.1684 [hep-th]].
}

\lref\maxdiegoone{
D.~Garc\'\i{}a Sep\'ulveda and M.~Guillen,
``A pure spinor twistor description of the $D = 10$ superparticle,''
JHEP {\bf 08}, 130 (2020).
[arXiv:2006.06023 [hep-th]].
}

\lref\maxdiegotwo{
D.~G.~Sep\'ulveda and M.~Guillen,
``A Pure Spinor Twistor Description of Ambitwistor Strings,''
[arXiv:2006.06025 [hep-th]].
%2 citations counted in INSPIRE as of 16 Dec 2022
}

\lref\nmmax{
N.~Berkovits, M.~Guillen, and L.~Mason,
``Supertwistor description of ambitwistor strings,''
JHEP {\bf 01}, 020 (2020).
[arXiv:1908.06899 [hep-th]].
}

\lref\maxmasoncasaliberkovits{
N.~Berkovits, E.~Casali, M.~Guillen, and L.~Mason,
``Notes on the $D=11$ pure spinor superparticle,''
JHEP {\bf 08}, 178 (2019).
[arXiv:1905.03737 [hep-th]].
}

\lref\maxthesis{
M.~Guillen,
``Pure spinors and $D=11$ supergravity,''
[arXiv:2006.06014 [hep-th]].
}

\lref\Berkovitsparticle{
N.~Berkovits, ``Covariant quantization of the superparticle using pure spinors'', \href{https://doi.org/10.1088/1126-6708/2001/09/016}{ {\it JHEP} {\bf 09} (2001) 016} [\href{https://arxiv.org/abs/hep-th/0105050}{{ hep-th/0105050}}].
}

\lref\BenShahar{
M.~Ben-Shahar and M.~Guillen, ``Superspace expansion of the 11D linearized superfields in the pure spinor formalism, and the covariant vertex operator'', \href{https://doi.org/10.1007/JHEP09(2023)018}{ {\it JHEP} {\bf 09} (2023) 018} [\href{https://arxiv.org/abs/2305.19898}{{ 2305.19898}}].
}

\lref\mafracohomology{
C.~R.~Mafra,
``Towards Field Theory Amplitudes From the Cohomology of Pure Spinor Superspace,''
doi.10.1007/JHEP11(2010)096.
}

\lref\Berkovitstopological{
N.~Berkovits,
``Pure Spinor Formalism as an N = 2 Topological String,'' JHEP
0510(2005) 089, hep-th/0509120
}

\lref\selivanov{
K.~G.~Selivanov,
`` On tree form-factors in (supersymmetric) Yang-Mills theory '' Commun.Math.Phys. 208 (2000) 671-687
}

\lref\Mafraalgorithm{
E.~Bridges and C.~R.~Mafra,
``Algorithmic construction of SYM multiparticle superfields in the BCJ gauge,'' JHEP 10 (2019) 022
}

\lref\ghostnumberzero{
M.~Guillen, M.~dos~Santos, and E.~Viana
``The 11D pure spinor ghost number zero vertex operator,'' to appear.
}

\lref\amplitudeeleven{
M.~Guillen, M.~dos~Santos, and E.~Viana,
``Tree-level 11D supergravity amplitudes from the pure spinor worldline,'' to appear.
}

\lref\mafraphd{
C.~Mafra,
``Superstring Scattering Amplitudes with the Pure Spinor Formalism'', PhD thesis, Sao Paulo IFT
}

\lref\maframultiparticles{
S.~Lee, C. R.~Mafra, and O.~Schlotterer
``Non-linear gauge transformations in D=10 SYM theory and the BCJ duality,'' JHEP 03 (2016) 090
}

\lref\wittentwistor{
E.~Witten,
``Twistor-Like Transform In Ten-Dimensions''
Nucl.Phys. B 266, 245 (1986).
}

\lref\Siegelsuperfields{
W.~Siegel,
``Superfields in Higher Dimensional Space-time''
Phys. Lett. B 80, 220
(1979).
}

\lref\Mizera{
S.~Mizera and B.~Skrzypek
``Perturbiner Methods for Effective Field Theories and the Double Copy''
JHEP 10 (2018) 018.
}

\lref\Siegelworldline{
P.~Dai, Y.~Huang, and W.~Siegel,
``Worldgraph approach to Yang-Mills amplitudes from N= 2 spinning particle''
Journal of High Energy Physics  (2008) 010
}

\lref\MafraNMoneloop{
C.~R.~Mafra and C.~Stahn,
``The One-loop Open Superstring Massless Five-point Amplitude
with the Non-Minimal Pure Spinor Formalism''
JHEP 03 (2009) 126
}

\lref\tamingbghos{
M.~Guillen
``Taming the 11D pure spinor b-ghost'',
JHEP 03 (2023) 135
}

\lref\BG{
F.~A.~Berends and W.~T.~Giele,
``Recursive Calculations for Processes with n Gluons'',
Nucl. Phys. B306 (1988) 759.
}

\lref\MafratowardsI{
C.~R.~Mafra and O.~Schlotterer,
``Towards the N-point one-loop superstring amplitude. Part I. Pure spinors and superfield kinematics,”
JHEP 08, 090 (2019). [arXiv:1812.10969 [hep-th]].
}

\lref\MafratowardsII{
C.~R.~Mafra and O.~Schlotterer,
``Towards the N-point one-loop superstring amplitude.
Part II. Worldsheet functions and their duality to kinematics,”
JHEP 08, 091 (2019). [arXiv:1812.10970 [hep-th]].
}

\lref\Mafratwoloops{
E.~D’Hoker, C.~R.~Mafra, B.~Pioline, and O.~Schlotterer,
``Two-loop superstring fivepoint amplitudes. Part I. Construction via chiral splitting and pure spinors,”
JHEP 08, 135 (2020). [arXiv:2006.05270 [hep-th]].
}

\lref\MafraGomes{
H.~Gomez and C.~R.~Mafra,
``The closed-string 3-loop amplitude and S-duality,”
JHEP 10, 217 (2013). [arXiv:1308.6567 [hep-th]].
}

\lref\Bjornsson{
J.~Bjornsson.
``Multi-loop amplitudes in maximally supersymmetric pure
spinor field theory,"
JHEP, 01:002, 2011.
}

\lref\Mafrarecursive{
C.~R.~Mafra, O.~Schlotterer, S.~Stieberger, and D.~Tsimpis,
``A recursive method for SYM n-point tree amplitudes'',
Phys. Rev. D 83 126012 (2011). [arXiv:1012.3981 [hep-th]].
}

\lref\Mafranonlinear{
C.~R.~Mafra and O.~Schlotterer,
``Solution to the nonlinear field equations of ten dimensional supersymmetric Yang-Mills theory,''
Phys. Rev. D 92 066001 (2015). [arXiv:1501.05562 [hep-th]].
}

\lref\Strassler{
M.~J.~Strassler.
``Field theory without Feynman diagrams: One-loop effective actions."
Nuclear Physics B, 385(1–2):145–184, October 1992.
}

\lref\Diana{
Y.~Du and D.~Vaman.
``Tree-level Graviton Scattering in the Worldline Formalism."
8 2023. 2308.11326 [hep-th]
}

\lref\Zvi{
Z.~Bern and D.~A.~Kosower.
``The Computation of loop amplitudes in gauge theories."
Nucl. Phys. B, 379:451–561, 1992.
}

\lref\Superfieldscederwall{
M.~Cederwall,
``Pure spinor superfields – an overview,”
Springer Proc. Phys. 153, 61-93 (2014). [arXiv:1307.1762 [hep-th]].
}

\lref\Superfieldscederwalleleven{
M.~Cederwall,
``D=11 supergravity with manifest supersymmetry,”
Mod. Phys. Lett. A 25, 3201-3212 (2010). [arXiv:1001.0112 [hep-th]].
}

\lref\ghostnumberzero{
M.~Guillen, M.~dos~Santos, and E.~Viana,
``The 11D pure spinor ghost number zero vertex operator,'' to appear.
}

\lref\equivalenceguillen{
M.~Guillen,
``Equivalence of the 11D pure spinor and Brink-Schwarz-like superparticle cohomologies,''
Phys. Rev. D 97, no.6, 066002 (2018). 
[arXiv:1705.06316 [hep-th]].
}

\lref\infiniteberkovits{
N.~Berkovits,
``Infinite Tension Limit of the Pure Spinor Superstring,''
JHEP 03, 017 (2014).
[arXiv:1311.4156 [hep-th]].
}

\lref\thalesrenann{
T.~Azevedo and R.~L.~Jusinskas,
``Connecting the ambitwistor and the sectorized heterotic strings,''
JHEP 10, 216 (2017).
[arXiv:1707.08840 [hep-th]].
}

\lref\guillenchiral{
M.~Guillen,
``Green-Schwarz and pure spinor formulations of chiral strings,''
JHEP 12, 029 (2021).
[arXiv:2108.11724 [hep-th]].
}

\lref\gomezyuan{
H.~Gomez and E.~Y.~Yuan,
``N-point tree-level scattering amplitude in the new Berkovits` string,''
JHEP 04, 046 (2014).
[arXiv:1312.5485 [hep-th]].
}

\lref\chandiavallilo{
O.~Chandia and B.~C.~Vallilo,
``Relating the b ghost and the vertex operators of the pure spinor superstring'',
JHEP 03 (2021) 165, [DOI: 10.1007/JHEP03(2021)165], [arXiv:2101.01129 [hep-th]]
}

\font\mbb=msbm10 
\newfam\bbb
\textfont\bbb=\mbb

\def\startcenter{%
  \par
  \begingroup
  \leftskip=0pt plus 1fil
  \rightskip=\leftskip
  \parindent=0pt
  \parfillskip=0pt
}
\def\stopcenter{%
  \par
  \endgroup
}

\listtoc
\writetoc
\filbreak

\newsec Introduction

The pure spinor formalism for superstrings \BerkovitsPS\ has proven to be the most powerful and efficient framework for calculating string scattering amplitudes, surpassing other standard formalisms such as the Ramond-Neveu-Schwarz (RNS) and Green-Schwarz (GS) \refs{\MafraNpoint, \MafratowardsI, \MafratowardsII, \Mafratwoloops, \MafraGomes}. Similarly, the 10D pure spinor superparticle, constructed in \Berkovitsparticle\ as the field theory limit of the pure spinor superstring, as well as its ambitwistorial counterpart \refs{\infiniteberkovits,\thalesrenann,\guillenchiral}, have been demonstrated to be highly effective for computing 10D super-Yang-Mills (SYM) amplitudes \refs{\Bjornsson, \maxmaor,\gomezyuan}. Furthermore, non-minimal pure spinor variables have been used to derive manifestly supersymmetric actions for maximally supersymmetric theories, including 10D SYM and 11D Supergravity \refs{\Superfieldscederwall, \Superfieldscederwalleleven}. Together, these developments highlight the pure spinor superparticle as a promising framework for computing field-theory scattering amplitudes.

\medskip
In string theory, amplitudes are typically formulated within a first-quantized framework, where we consider the quantum theory of a single string propagating through spacetime, and interactions are represented by insertions on the worldsheet of this string. In contrast, quantum field theory (QFT) is usually developed within a second-quantized approach, where particles are treated as excitations of a quantum field, and the individual trajectories of particles are not considered. However, the amplitudes of QFT can also be reformulated in a first-quantized manner, as illustrated by Strassler through the worldline formalism in \Strassler, firstly introduced for one-loop computations and partially extended to treat tree-level amplitudes in \refs{\Siegelworldline,\Diana}. This approach utilizes the path integral of a single particle, with interaction insertions representing the effects of background fields. The worldline formalism is particularly advantageous in scenarios where standard perturbative QFT techniques are unavailable or cumbersome, such as in the case of gravity. In particular, the worldline formalism is an effective tool for studying scattering amplitudes in 10D SYM theory and 11D Supergravity, as these theories can be expressed in a first-quantized framework that manifests supersymmetry.

\medskip
In this paper, we will use the worldline formalism to compute tree-level scattering amplitudes of 10D SYM using the pure spinor formulation of the superparticle, which exhibits manifest super-Poincar\'e covariance \Berkovitsparticle. The use of the worldline formalism for such calculations is challenging, mainly due to the structure of the contractions and operators associated with the worldline \Siegelworldline. In this work, we aim to explore these structures within the framework of a BRST-quantized theory. A key aspect of our investigation is the insertion of pinching operators---interpreted as arising from higher-order corrections in the external fields--—which are crucial for the BRST invariance of the amplitude. We demonstrate that these operators correspond to the higher-order poles in the OPE between two integrated vertices in string theory, $U(z)U(0)$, and have the role of contact terms in the amplitude, enabling us to express the 10D SYM amplitude in terms of cubic kinematic numerators that explicitly satisfy the generalized Jacobi identities.

\medskip
The computation of scattering amplitudes within the pure spinor superparticle framework was addressed for the first time in \Bjornsson, using canonical quantization. In this work, we will perform a similar construction, but using the path integral over the worldline instead of canonical quantization. In particular, we will explicitly produce all the necessary insertions on the worldline, including subtree diagrams, and interpret these as coming from BRST invariance direction on the action. 

In \MafraNpoint, the tree-level N-point amplitude of 10D SYM was computed by taking the infinite tension limit of the disk open superstring amplitude. A key contribution of this paper was the identification of the BRST cohomology structure of the kinematical numerators, known as the BRST building blocks, which allow the amplitude to be bootstrapped using BRST invariance arguments. These building blocks naturally emerge from the superstring computation, and in our approach, we show that the same structure appears within the worldline formalism. Remarkably, the ideas presented in this work can be appropriately generalized to 11D for computing tree-level supergravity scattering amplitudes. We discuss these issues further in the accompanying papers \refs{\amplitudeeleven,\ghostnumberzero}.

\medskip
This paper is organized as follows. In section 2, we will review the 10D pure spinor superparticle and its BRST cohomology. The non-minimal pure spinors are introduced to construct the $b-$ghost, satisfying $\{Q,b\}=\half P^2$. In section (2.3), we will couple the pure spinor superparticle to a SYM background, as was done in \Berkovitsparticle. 
The insertions needed for the worldline tree amplitude will arise as the multiparticle expansion
of the on-shell background fields, and their defining properties will be seen as deriving from
the BRST consistency of the interacting action. In this section, the insertions needed
for the worldline will be seen to satisfy the same master equations as the Berends-Giele currents
of SYM. In section 3, we will provide a prescription 
for the 10D SYM tree amplitude in the pure spinor worldline formalism and prove the BRST 
invariance of the prescription. In sections (3.3), (3.4), and (3.5) we will explicitly 
compute the 4-point, 5-point, and N-point amplitudes of the 10D SYM, showing that our result 
agrees with the amplitudes obtained as the low-energy limit of the open superstring 
in \refs{\MafraNpoint,\Mafrarecursive,\Mafranonlinear}.

\seclab\secone

\noindent  

\newsec Pure spinor superparticle in 10D

\seclab\sectwo

\noindent The 10D pure spinor superparticle action in an Euclidean flat background is given by \Berkovitsparticle
\eqnn \tendaction
$$ \eqalignno{
S &= \int d\tau \left(\dot{x}^m p_m - \half p^2 + p_{\alpha} \dot{\theta}^{\alpha}  + \omega_{\alpha} \dot{\lambda}^{\alpha}\right), & 
\tendaction
}
$$
where $x^m$ is $SO(10)$ vector with $m = 1, \dots, 10$, $\theta^\alpha$ is a fermionic Majorana-Weyl spinor and $\lambda^\alpha$ is a bosonic Weyl spinor, with $\alpha=1,\cdots,16$. The fields $p_m$, $ p_\alpha$, $w_\alpha$ are the conjugate momenta corresponding to $x^m$, $\theta^\alpha$, $\lambda^\alpha$ respectively. The spinor fields $\lambda^{\alpha}$ satisfy the pure spinor constraint
\eqnn \purespinorconstraint
$$ \eqalignno{
\lambda^{\alpha} \gamma^m_{\alpha \beta} \lambda^{\beta} &= 0, & 
\purespinorconstraint
}
$$
where the $\gamma^m_{\alpha\beta}$ and $\gamma_m^{\alpha \beta}$ are the $16 \times 16$ symmetric Pauli matrices in 10D, which satisfy $(\gamma^{(m})^{\alpha\delta}\gamma^{n)}_{\delta\beta}= 2 \eta^{mn} \delta^\alpha_\beta$. The constraints \purespinorconstraint\ generate the pure spinor gauge transformation
\eqnn \purespinorgaugetransformation
$$ \eqalignno{
\delta \omega_{\alpha} &= \Lambda_m (\gamma^m \lambda)_{\alpha}, & 
\purespinorgaugetransformation
}
$$
which leaves the action \tendaction\ invariant for arbitrary $\Lambda_m$. The constraints \purespinorconstraint\ and gauge symmetry \purespinorgaugetransformation\ reduce the $16$ complex components of each bosonic ghost to $11$. The linear combinations of the $\omega_{\alpha}$ field that are invariant under the pure spinor gauge transformations are the ghost charge current $J$ and Lorentz current $N^{mn}$, given by
\eqnn \tendJandN
$$ \eqalignno{
J &= -\omega_{\alpha} \lambda^{\alpha}, \quad N_{mn} = \half (\omega \gamma_{mn} \lambda). & 
\tendJandN
}
$$
Under the action of $J$, $\lambda^{\alpha}$ has ghost number $+1$ and $\omega_{\alpha}$ has ghost number $-1$. 
The superparticle action \tendaction\ implies the following Euclidean canonical commutators
\eqnn\ccr
$$\eqalignno{
    [p_m, x^n] & = -\delta_{m}^n, \quad \{ p_{\alpha}, \theta^{\beta}\} = \delta_{\alpha}^{\beta}, \quad [\omega_{\alpha}, \lambda^{\beta}] =  -\delta_{\alpha}^{\beta}. &\ccr
}
$$
To be more precise, the Poisson brackets and commutators associated with the bosonic ghosts have to be defined in a way that preserves the pure spinor constraint, $[\omega_{\alpha}, (\lambda \gamma^m \lambda)] = 0$. This implies that $\{\omega_{\alpha}, \lambda^{\beta}\}$ has to be corrected by a projection operator $K_{\alpha}^{\beta}$. For the computations carried over in this work, the existence of $K_{\alpha}^{\beta}$ has no consequence.

We will be interested in computing path integral correlation functions for the superparticle action. It is convenient to write the operator algebra obtained by quantization of \tendaction\ in terms of the path integral fields. We obtain the following free field contractions
\eqnn\wick
$$\eqalignno{
p_m(\tau_1) x^n(\tau_2) & \sim - \delta_m^{n} \sigma_{12}, & \cr
p_{\alpha}(\tau_1) \theta^{\beta}(\tau_2) & \sim \delta_{\alpha}^{\beta} \sigma_{12} , & \cr
\omega_{\alpha}(\tau_1) \lambda^{\beta}(\tau_2) & \sim - \delta_{\alpha}^{\beta}\sigma_{12}, &\wick
}
$$
where we introduced the convenient notation $\half$sign$(\tau_i -\tau_j) = \sigma_{ij}$. These contractions are equivalent to the canonical commutation relations. This can be checked by using the map between graded commutators and operator insertions on the path integral, given by
\eqnn\mapcommutatorspathintegral
$$\eqalignno{
[A(\tau), B(\tau)\} & \sim A(\tau+\epsilon)B(\tau) \mp B(\tau)A(\tau-\epsilon), &\mapcommutatorspathintegral
}
$$
where $\epsilon \to 0$ and the sign is $-$ for commutators, and $+$ for anticommutators.

The superparticle action is complemented by the BRST operator
\eqnn\BRST
$$\eqalignno{
&Q = \lambda^\alpha d_\alpha,&\BRST
}$$
where $d_\alpha = p_\alpha - {1\over2}(\gamma^m\theta)_\alpha p_m$ are the fermionic constraints of the Brink-Schwarz superparticle \brinkschwarz. The physical spectrum is defined as the cohomology of the BRST operator \BRST. To find the physical spectrum of the theory, it is convenient to split the BRST cohomology into different ghost number sectors. It can be proven that the cohomology is trivial for ghost numbers larger than 3, so an arbitrary wave function on the cohomology can be expanded as
\eqnn\wavefunction
$$\eqalignno{
\Psi(x, \theta, \lambda) &= \Psi^{(0)} + \Psi^{(1)} + \Psi^{(2)} + \Psi^{(3)}.&\wavefunction
}$$
where $\Psi^{(k)}$ has ghost number $k$. The wave functions on the ghost number sectors $0$, $1$, $2$, and $3$ describe, respectively, the ghosts, fields, antifields, and antighosts of the BV formulation of 10D SYM  \Berkovitsparticle, so the pure spinor superparticle furnishes a description of 10 SYM. The ghost number 1 part is the most relevant for our purposes since it corresponds to the physical degrees of freedom of 10D SYM. To see this, one can write $\Psi^{(1)} = \lambda^{\alpha} A_{\alpha}$ where $A^{\alpha}$ is an arbitrary wave function of the matter variables, $A^{\alpha} = A^{\alpha}(x, \theta)$. The BRST cohomology implies equations for this wave function
\eqnn\conditions
$$\eqalignno{
&Q\Psi^{(1)} = 0,\ \ \ \ \delta\Psi^{(1)} = Q\Lambda,&\conditions
}$$
that we recognize respectively as the linearized equations of motion and gauge symmetry of 10D SYM \Mafraalgorithm
\eqnn\SYMeq
$$\eqalignno{
&(\gamma^{mnpqr})^{\alpha\beta}D_\alpha A_\beta = 0,\ \ \ \ \delta A_\alpha = D_\alpha\Lambda,&\SYMeq
}$$
where $D_\alpha = \partial_\alpha + \half(\gamma^m\theta)_\alpha\partial_m$ is the covariant superderivative.

In the usual BRST quantization of theories carrying diffeomorphism invariance, a $b-$ghost operator exists satisfying $\{Q, b\} = \half p^2$. This is useful because it allows one to trivialize the BRST cohomology for off-shell fields (fields carrying a non-zero eigenvalue of $p^2$). Even though the pure spinor superparticle \tendaction\ is a BRST gauge-fixed action, this theory has no obvious analogous of the $b$-ghost operator because such an object must carry ghost number $-1$. This motivates the enlargement of the action \tendaction\ through the introduction of a new set of variables, called the non-minimal pure spinor variables, which allows one to define a composite $b-$ghost operator.

\subsec Non-minimal variables

It was shown in \Berkovitstopological\ that one can define a composite $b$-ghost operator for the pure spinor string by adding non-minimal variables to the pure spinor formalism. The same thing can be done for the pure spinor particle. The set of new variables consists of a bosonic pure spinor $\bar\lambda_\alpha$ and a fermionic spinor $r_\alpha$ satisfying the constraints
\eqnn\nonminimal
$$\eqalignno{
&\bar\lambda\gamma^m\bar\lambda = 0, \ \ \ \ \ \bar\lambda\gamma^mr = 0.&\nonminimal
}$$
These non-minimal variables $\bar\lambda_{\alpha}$ and $r^{\alpha}$ enter into the enlarged action together with their respective conjugate momenta $\bar\omega^{\alpha}$ and $s^{\alpha}$
\eqnn\nonminimalaction
$$\eqalignno{
S &= \int d\tau \left(\dot{x}^m p_m - \half p^2 + p_{\alpha} \dot{\theta}^{\alpha}  + \omega_{\alpha} \dot{\lambda}^{\alpha} +  \bar\omega^{\alpha} \dot{\bar\lambda}_{\alpha} + s^\alpha\dot r_\alpha\right). & 
\nonminimalaction
}$$
To preserve the BRST cohomology, in such a way that this new set of variables describes the same physics as the action studied in the last subsection, the BRST charge must be modified according to the quartet argument \Berkovitstopological
\eqnn\nonminimalBRST
$$\eqalignno{
&Q = \lambda^\alpha d_\alpha + r_\alpha\bar\omega^\alpha.&\nonminimalBRST
}$$
Using the non-minimal variables, one can construct the composite $b$-ghost satisfying
\eqnn\Qb
$$\eqalignno{
&\{Q,b\} = {p^2\over2}. &\Qb
}$$
The explicit form of the $b$-ghost was initially proposed in \Berkovitstopological\ and is a complicated expression. The expression for the $b$-ghost has been disentangled in \dynamical\ and a further simplification was proposed in \maxmaor\ using the so-called physical operators, introduced in \psborninfeld. Throughout this paper, we will use this version for the $b$-ghost, which reads
\eqnn\btend
$$\eqalignno{
& b = - {\bf \Delta}_m {\bf A}^m, & \btend
}$$
where the bold letters denote the so-called physical operators defined in \maxmaor\ and \psborninfeld, given by
\eqnn \physicaloperatorAm
\eqnn \physicaloperatorDelta
$$\eqalignno{
{\bf{A}}_{m} & = {(\bar{\lambda}\gamma_{m}d)\over 2(\lambda\bar{\lambda})} + {(\bar{\lambda}\gamma_{mnp}r)\over 8(\lambda\bar{\lambda})^{2}}N^{np}, & \physicaloperatorAm \cr
\bf{\Delta}_{m} &= -p_{m} - {(\lambda\gamma^{mn}r)\over 4(\lambda\bar{\lambda})}\bf{A}_n. & \physicaloperatorDelta
}$$

\subsec{Uniqueness in the BRST cohomology}

It proves useful to define another operator from the $b$-ghost field. 
This is obtained by promoting the canonical momenta in $b$ 
to first-order differential operators, that is
\eqnn \momentumoperatormap
$$ \eqalignno{
 & p_m \to - \partial_m, \quad d_{\alpha} \to D_{\alpha}, \quad \omega_{\alpha} \to - \partial_{\lambda^{\alpha}}, & \momentumoperatormap
}$$
as suggested by equation \ccr. Doing this to the canonical momenta appearing in equation \btend\ defines the operator $b_0$. It can be thought of as the zero-mode of the
Laurent expansion of the string $b$-ghost, $(b(z) = b_0/z^2 + \cdots)$, but for the particle. In particular, equation \Qb\ implies that the $b_0$ operator satisfies the BRST cohomology
\eqnn \Qbzero
$$ \eqalignno{
& \{Q, b_0\} = \Box, & \Qbzero
}$$
where $\Box$ is the Laplacian $\partial^m \partial_m$. This equation trivializes the BRST cohomology in the space orthogonal to the kernel of the Laplacian. Indeed, suppose that $\Phi$ is a BRST closed wave function with non-zero energy, i.e. $\{Q, \Phi\} = 0$ and $\Box \Phi = h \Phi$ for $h \neq 0$. These conditions imply that
\eqnn \BRSTuniqueness
$$ \eqalignno{
& \Phi = \left\{Q, { b_0(\Phi) \over h } \right\}, & \BRSTuniqueness
}$$
which establishes that every BRST closed wave function with non-zero energy is also BRST exact. As we will see, this is enough to reformulate the $N$-point tree-level amplitude as a question on BRST cohomology.

\subsec Superparticle in a SYM background

The worldline prescription depends on the coupling of the particle to an external background. We will perform this construction in this subsection, and we shall see that BRST symmetry implies that the external fields satisfy the equations of motion of multiparticle SYM \Mafraalgorithm.

The pure spinor superparticle action \tendaction\ can be coupled to a SYM background by introducing external fields. To describe the color degrees of freedom of SYM, one must also introduce fermions on the worldline. For $SO(N)$ (or $U(N))$ gauge group, one needs $N$ real (or complex) fermions. For instance, for $SO(N)$, the Euclidean action is \Berkovitsparticle
\eqnn\actioncoupled
$$\eqalignno{
S^{coupled} & =  \int d\tau \Big\{ \dot{x}^m\tilde p_m - \half\tilde p^2 + \dot{\theta}^{\alpha}\tilde p_{\alpha} + \dot{\lambda}^{\alpha} \omega_{\alpha} \cr
& - {1\over 2} \eta_I \dot{\eta}^I - g \eta_{I} \eta_{J} \Big( \dot{\theta}^{\alpha} \bA_{\alpha}^{IJ} + \Pi^m \bA^{IJ}_m + \tilde d_{\alpha} \bW^{IJ \alpha } + \half N^{mn} \bF^{IJ}_{mn} \Big) \Big\},&\actioncoupled
}$$
\noindent where $\Pi^m = \dot{x}^m + {1\over2} (\dot{\theta} \gamma^m \theta)$, and $\eta_I$ are worldline fermions that encode the color degrees of freedom. For $SO(N)$, the indices $I$ and $J$ are in the fundamental representation, and the fields $\bA_{\alpha}^{IJ}, \bA^{IJ}_m, \bW^{IJ\alpha}, \bF_{mn}^{IJ}$ depend on $x^m$ and $\theta^{\alpha}$ and should be understood as valued in the Lie-algebra. For $SU(N)$, the indices $I$ and $J$ belong to the fundamental and antifundamental, as appropriate. 

The external fields describe multiparticle SYM \mafraoli, as we will see shortly. It is convenient to use a basis $T_a^{IJ}$ for the Lie-algebra and redefine the external fields in this basis, e.g. $\eta_{I} \eta_{J} \bA_{\alpha}^{IJ} = \eta_{I} \eta_{J} T_a^{IJ} \bA_{\alpha}^a$. Defining further $T_a = \eta_{I} \eta_{J} T_{a}^{IJ}$, it is clear that $[T_a, T_b] = if_{ab}^c T_c$ if we normalize the internal fermions $\eta_{I}$ appropriately, where $f_{ab}^c$ are the structure constants of the Lie-algebra. We will omit the Lie-algebra generators and write simply $\bA_{\alpha} \equiv T_a \bA_{\alpha}^{a}$. 
We will see that the superfields $\bA_{\alpha}, \bA_m, \bW^{\alpha}$ and $\bF_{mn}$ satisfy a perturbiner expansion where each coefficient is a Berends-Giele current. Performing this change of basis, the action reads
\eqnn\actioncoupledsimp
$$\eqalignno{
S^{coupled} & =  \int d\tau \Big\{ \dot{x}^m\tilde p_m - \half\tilde p^2 + \dot{\theta}^{\alpha}\tilde p_{\alpha} + \dot{\lambda}^{\alpha} \omega_{\alpha} \cr
& - {1\over 2} \eta_I \dot{\eta}^I - g \Big( \dot{\theta}^{\alpha} \bA_{\alpha} + \Pi^m \bA_m + \tilde d_{\alpha} \bW^{\alpha } + \half N^{mn} \bF_{mn} \Big) \Big\}.&\actioncoupledsimp
}$$
As happens for the free particle, this pure spinor action is complemented by a BRST charge, which is given by
\eqnn\BRSTcoupled
$$\eqalignno{
& Q^{coupled} = \lambda^\alpha\tilde d_\alpha. &\BRSTcoupled
}$$
Moreover, the space of physical states is once again identified with the cohomology of the BRST charge.

The fields $\tilde p_m$ and $\tilde p_\alpha$ appearing in the action \actioncoupledsimp\ and BRST charge \BRSTcoupled\ are not equal to the canonical momenta of the corresponding variables. Taking derivatives with respect to the velocities, we can compute the canonical momenta to be
\eqnn\momenta
$$\eqalignno{
{p}_m & = {\partial L\over\partial \dot{x}^m} = \tilde p_m - g \bA_{m}, \cr
{p}_{\alpha} & = {\partial L\over\partial \dot{\theta}^{\alpha}} = \tilde p_{\alpha} - g \left( \bA_{\alpha} + {1\over2} (\gamma^m \theta)_{\alpha} \bA_m \right).
}$$
Further insight is obtained if we transform the action \actioncoupledsimp\ to express it in Hamiltonian form by performing a Legendre transformation on the velocities. The Hamiltonian is given by 
\eqnn\Hamiltoninan
$$\eqalignno{
&H^{coupled}(x^m, {p}_m, \theta^{\alpha}, {p}_{\alpha}, \lambda^{\alpha}, \omega_{\alpha}) = (\dot{x}^m {p}_m + \dot{\theta}^{\alpha} {p}_{\alpha} + \dot{\lambda}^{\alpha} \omega_{\alpha} - L) \Big|_{\dot{q} = \dot{q}(p)}\cr
& = \half {p}^2 + g \left({p}^m \bA_m + {d}_{\alpha} \bW^{\alpha} + \half N^{mn} \bF_{mn} \right) + g^2 \left(\half \bA^2 + \bA_{\alpha} \bW^{\alpha} \right), & \Hamiltoninan
}$$
where $d_\alpha = p_\alpha - \half (\gamma^m\theta)_\alpha p_m$. The first term of 
the Hamiltonian is the energy of the free superparticle, $H = \half p^2$, and the terms 
proportional to $g$ are the perturbation due to the SYM background. 
We write $H^{coupled} = H + \bU$, splitting the free and interacting pieces of the 
Hamiltonian. The object $\bU$ will turn out to contain the correct 
insertions for the worldline amplitude, and plays a similar role to the vertex operators of 
usual worldsheet string theory. However, as we will see, to describe the full tree-level 
amplitude, we must allow $\bU$ to be non-local. Using equation \Hamiltoninan, we see that $\bU$ can be expanded as
\eqnn\U
$$\eqalignno{
&\bU = g \left({p}^m \bA_m + {d}_{\alpha} \bW^{\alpha} + \half N^{mn} \bF_{mn} \right) + g^2 \left(\half \bA^2 + \bA_{\alpha} \bW^{\alpha} \right).&\U
}$$
The Hamiltonian form of the action is given by
\eqnn\actionH
$$\eqalignno{
&S^{coupled} = S - \int d\tau\bU,&\actionH
}$$
where $S = \int d\tau \left( \dot{x}^m {p}_m + \dot{\theta}^{\alpha} {p}_{\alpha} + \dot{\lambda}^{\alpha} \omega_{\alpha} - {1\over2} p^2 \right)$ is the free superparticle action. It is also convenient to express the BRST charge $ Q^{coupled} $ in terms of the canonical momenta. It splits into the free BRST charge $Q = \lambda^{\alpha} d_{\alpha}$ and a perturbation term
\eqnn\BRSTchargecoupled
$$\eqalignno{
&Q^{coupled} = \lambda^{\alpha}\tilde d_{\alpha} = \lambda^{\alpha} {d}_{\alpha} + g \lambda^{\alpha} \bA_{\alpha} = Q + \bV,&\BRSTchargecoupled
}$$
where $\bV$ is defined as
\eqnn\multiparticleV
$$\eqalignno{
    & \bV = g \lambda^{\alpha} \bA_{\alpha}. &\multiparticleV
}$$
The objects $\bU$ and $\bV$ will turn out to be the generating functions for the 
Berends-Giele currents of the theory.

The theory is only consistent when the background fields obey the BRST consistency conditions, as follows. The BRST charge must be nilpotent, $\{Q^{coupled},Q^{coupled}\}=0$. This implies
\eqnn\consistenceBRSTV
$$\eqalignno{
& \{Q,\bV\}=-\half \{\bV,\bV\}. &\consistenceBRSTV
}$$
Further, the BRST charge must be conserved in time, $[H^{coupled}, Q^{coupled}] = 0$. Expanding in terms of the background vertices, we find
\eqnn\consistenceBRSTU
$$\eqalignno{
&[Q, \bU] = [H,\bV] + [\bU, \bV].&\consistenceBRSTU
}$$
These equations are equivalent to the SYM equations of motion for the external SYM fields. Indeed, expanding $\bU$ as in \U\ and $\bV$ as in \multiparticleV, the equation that follows from \consistenceBRSTV\ is
\eqnn\eqA
$$\eqalignno{
& (\gamma^{mnpqr})^{\alpha\beta}\left(D_\alpha\bA_\beta + \{\bA_\alpha,\bA_\beta\}\right)=0,&\eqA
}
$$
while \consistenceBRSTU\ implies the equations
\eqnn\eqmotUone
\eqnn\eqmotUtwo
\eqnn\eqmotUthree
$$\eqalignno{
[\nabla_\alpha,\bA_m] &= [\partial_m,\bA_\alpha] + (\gamma_m\bW)_\alpha,&\eqmotUone\cr
\{\nabla_\alpha,\bW^\beta\} &= {1\over4}(\gamma^{mn})_\alpha^{\ \beta}\bF_{mn}, &\eqmotUtwo\cr
[\nabla_\alpha,\bF_{mn}] &= \nabla_{[m}(\gamma_{n]}\bW)_\alpha, &\eqmotUthree
}$$
\noindent where we defined the covariant derivatives $\nabla_\alpha = D_\alpha + \bA_\alpha$, with $D_\alpha = \partial_\alpha +\half (\theta\gamma^m)_\alpha\partial_m$, and $\nabla_m = \partial_m + \bA_m $. These imply the equations of motion of 10D non-Abelian SYM \refs{\mafraoli,\Mafraalgorithm,\wittentwistor,\Siegelsuperfields}.

These equations can be solved order by order on the SYM coupling $g$ using the perturbiner method of Selivanov \refs{\selivanov,\Mafraalgorithm,\Mizera}. In this approach, one expands the fields in terms of multiparticle asymptotic states. Each asymptotic particle $i$ is proportional to a plane wave $e^{i k_i \cdot x}$ and a color factor $T^i$. 

To organize the expansion, we define words $P$ containing letters from the natural numbers (eg. $P = 1253$), that might also stand for bracketing of words (eg. $P = [1,4] = 14 - 41$) \Mafraalgorithm. If $P$ is a word containing the letters $P_1 \dots P_m$, then we define the corresponding momentum $k_P = k_{P_1} + \dots + k_{P_m}$. The Lie-algebra generator with a word as superscript, $T^P$, is defined as $T^{P} = T^{P_1} \dots T^{P_m}$. Each natural number is used as the index of an external state. We expand the superfield $\bA^{\alpha}$ in terms of plane wave states
\eqnn\expansionAalpha
$$\eqalignno{
\bA_{\alpha} & = \sum_P {\cal A}_{\alpha P} T^P e^{k_Px} = {\cal A}_{\alpha i} T^i e^{k_ix} + {\cal A}_{\alpha ij} T^{ij} e^{k_{ij}x_{ij}} + \dots,&\expansionAalpha\cr
}$$
and similarly for $\bA_m,$ $\bW^{\alpha}$ and $\bF^{mn}$. The coefficients in the expansion are the Berends-Giele currents of SYM. We define the rank of an object with subscript $P$ as $|P|$, where $|P|$ is the number of letters in the word $P$. The rank is also the number of particles in the respective state (e.g. ${\cal A}_{\alpha 12}$ has rank $2$ and represents two-particle asymptotic states of SYM). 

Collecting the non-Abelian SYM equations \eqA-\eqmotUthree\ order by order in the coupling $g$ and the Lie-algebra generators $T^P$, one finds the equations that each of the rank-$|P|$ coefficients must satisfy. From now onwards, we also fix $g = 1$ after performing the expansion, to simplify the notation. The solution depends on the gauge choice. A convenient choice is the Lorenz gauge ($[\partial_m, \bA^m] = 0$), where the coefficients can be determined recursively. For ${\cal A}_{\alpha}^{P}$, for instance, one finds
\eqnn\AP
$$\eqalignno{
&{\cal A}_\alpha^P = {1\over s_P} \sum_{RQ=P}{\cal A}_{\alpha}^{[RQ]}, &\AP
}$$
where
\eqnn\recursionA
$$\eqalignno{
{ \cal A}_{\alpha}^{[PQ]} & = - {1\over2} \big[ {\cal A}_{\alpha}^P (k^P \cdot {\cal A}^Q) + {\cal A}^P_m (\gamma^m {\cal W}^P)_{\alpha} - (P \leftrightarrow Q) \big]&\recursionA\cr
}$$
and $s_P = \half k_P^2$. The notation $RQ = P$ refers to the deconcatenation sum of the word $P$---we sum over all {\it non-empty} words $R$ and $Q$ such that the union of the letters of $R$ and $Q$, ordered such that $R$ is to the left, gives $P$. For instance, if $P = 123$, then the tuple $(R,Q)$ takes the two values $(1, 23)$ and $(12, 3)$. The full solution for the Berrends-Giele currents in this gauge can be found in \maframultiparticles. To achieve our goal, all we need to know is that the Berrends-Giele constructed using the recursion \recursionA\ (and similarly for the remaining fields) are such that the action \actionH\ is BRST consistent in the sense of equations \consistenceBRSTU\ and \consistenceBRSTV.

There is a point we must stress here. Since the currents are non-local (due to the factors of $1/s_P$), this multiparticle action will also carry non-local information, which is ultimately what allows one to describe subtree diagrams in this worldline formalism.

We now turn to study general properties of the operators defined in \U\ 
and \multiparticleV, without direct reference to their SYM components. 
The perturbiner expansion of the SYM fields implies perturbiner expansions
for $\bU$ and $\bV$. In the latter case, we find
\eqnn\expansionV
$$\eqalignno{
\bV & = \sum_P V_P T^P = V_i T^i + V_{ij} T^{ij} + \dots.&\expansionV
}$$
The $V_P$ are Berends-Giele currents, which we will also refer to as unintegrated vertex currents. For $\bU$, the second term in 
equation $\U$ is of order four in the 
internal fermions of the worldline, which means that we must expand 
(e.g. for SU(N) group)
\eqnn\expansionU
$$\eqalignno{
\bU & = \sum_P U_P T^P e^{k_Px} + \sum_{PQ} D_{PQ} (T^P\odot T^Q)e^{k_{PQ}x},  &\expansionU
}$$
where $\odot$ is the symmetric tensor product. $U_P$ are the integrated vertex
currents, and $D_P$ will be refered to as contact terms or {\it pinching operators}.
We say that an object with $P$ as a subscript is of rank-$|P|$, where
$|P|$ is the number of letters in the word. 
The operators with a single index ($U_i$ and $V_i$) represent single-particle 
states. The operators with two anti-symmetric indices 
($U_{ij}$ and $V_{ij}$) are two-particle states.
Note that the coefficients $V_P$ and $U_P$ 
are non-local, since they can be written in terms of the 
SYM Berends-Giele currents defined previously. 
$D_P$ generally has less propagators than the other objects, which
justifies its interpretation as a contact term.

The second-order vertices represent quadratic effects equivalent to interaction with states containing two particles. In the string, we only need the rank-$1$ operators and can drop all of the higher-rank terms, due
to conformal symmetry on the worldsheet. This is not possible for the particle. The amplitude is only BRST invariant after adding all of the higher-rank vertices, as we will see.

The equations obeyed by the vertices at each rank are obtained by matching powers of the SYM coupling and the Lie-algebra generators in \consistenceBRSTV\ and \consistenceBRSTU,
as we did for the SYM fields. Plugging in equation \expansionV\ in equation \consistenceBRSTV, one finds the following equations for the unintegrated vertex up to quadratic order (the exponential factors are implicit)
\eqnn\rankoneV
\eqnn\ranktwoV
$$\eqalignno{
&\{Q,V_i\} = 0,&\rankoneV\cr
&\{Q,V_{ij}\} = -V_iV_j.&\ranktwoV
}$$
Plugging \expansionU\ into \consistenceBRSTU, one finds that up to quadratic order, the integrated vertices satisfy
\eqnn\rankoneU
\eqnn\ranktwoU
\eqnn\ranktwoD
$$\eqalignno{
& [Q,U_i] = [H,V_i], &\rankoneU\cr
& [Q,U_{ij}] = [H,V_{ij}] + U_iV_j - U_jV_i, &\ranktwoU\cr
& [Q,D_{ij}] = \half\left( [U_i,V_j] + [U_j,V_i] \right). &\ranktwoD
}$$
The linear equations \rankoneV\ and \rankoneU\ are the usual equations of motion of the vertices of pure spinor SYM, while the equations for $V_{ij}$ and $U_{ij}$ give the two-particle SYM equations of motion \refs{\mafraoli,\Mafraalgorithm}.  More generally, the equations obtained by the multiparticle vertices are
\eqnn\allrankV
\eqnn\allrankU
$$\eqalignno{
&\{Q,V_P\} = - \sum_{R+Q=P}V_RV_Q,&\allrankV\cr
& [Q,U_P] = [H,V_P] + \sum_{RQ=P}U_RV_Q - U_QV_R,&\allrankU
}$$
while the pinching operator satisfies
\eqnn\allrankD
$$\eqalignno{
& [Q,D_P] =  \half \sum_{RQ=P}[U_R,V_Q] + [U_Q,V_R]. &\allrankD
}$$
The deconcatenation sum $RQ = P$ was defined below equation \recursionA. Note that \allrankV\ is exactly the equation satisfied by Berends-Giele currents.

\subsec Matching to string theory OPEs

There is a match between these operators and the objects that one 
encounters in the string amplitude, which we will now review. 
Remember that the basic objects of pure spinor string theory are 
the vertex operators $U^s(z)$ (conformal weight one and ghost number zero)
and $V^s(z)$ (conformal weight zero and ghost number one). Both carry plane wave
factors $e^{k \cdot X(z)}$ (in our conventions $k$ is purely imaginary), which
produce the Koba-Nielsen factor. 
In this subsection we use the superscript $s$ to differentiate the string vertex operators from the objects 
we introduced before. We employ the OPE conventions of \MafraNpoint. 
The string vertex operators are constrained to satisfy the BRST cohomology
\eqnn\stringcohomology
$$\eqalignno{
& \{Q,V^s\} =  0, \quad \{Q, U^s\} = \partial V^s. &\stringcohomology
}$$
One should notice the similarity between \stringcohomology\ and the equations 
obeyed by the single-particle vertices of the particle, 
\rankoneV\ and \rankoneU. Indeed, there is a map between the vertex
operators of the string (and the OPEs constructed from these objects) and the 
particle vertex currents introduced above, that we now describe.
We will focus on the rank-two objects, which are simple to study.

The OPE between two integrated vertices in string theory is given by (we suppose
that $z_2 > z_1$ and factor out the contribution to the Koba-Nielsen factor) 
\eqnn\stringtheoryOPE
$$\eqalignno{
& U_1^s(z_1) U_2^s(z_2) \sim (z_2 - z_1)^{-k_1 \cdot k_2} \left({2(1 + s_{12}) \over  z_{12}^2} D^s_{12}(z_2) + { 1 \over z_{12}} U^s_{12}(z_2)\right). &\stringtheoryOPE
}$$
where $D^s_{12}$ is the double pole and $U^s_{12}(z_2)$ is the single pole.
We have factorized $2(1+s_{12})$ from the double pole for later convenience.
The $U^s_{12}$ operator represents a two-particle state coupling to the worldsheet,
so we imagine it should be related to the vertex current $U_{12}$ defined above. This is the
right intuition.

To derive the equations for $[Q, D_{12}^s]$ and $[Q, U_{12}^s]$, we will
compute the action of the BRST charge around an insertion containing $U_1^s(z_1)$
and $U_2^s(z_2)$. There are two ways we can perform the computation---by either 
acting with $Q$ first and then taking the OPE, or first taking the OPE and then 
acting with $Q$. 

In the first case, after acting with $Q$ on $U_1^s U_2^s$ we obtain $\partial_1 V_1^s(z_1) U_2^s(z_2) + U_1^s(z_1) \partial_2 V^s_2(z_2)$.
The OPE of this last expression might be computed with ease if we first introduce the multiparticle field $L_{12}$. This is
the singular piece of the $U^sV^s$ OPE (in this subsection, singular means more singular than the Koba-Nielsen factor), defined such that
\eqnn\defnL
$$\eqalignno{
    & U^s_1(z_1) V^s_2(z_2) = (z_2 - z_1)^{-k_1 \cdot k_2} \left(\frac{1}{z_{12}} L_{12}(z_1, z_2)\,\, + :U_1^s(z_1)V_2^s(z_2):\right) \cr
    & \sim  (z_2 - z_1)^{-k_1 \cdot k_2} \left(\frac{1}{z_{12}} L_{12}(z_2, z_2)+ \partial_1 L_{12}(z_1, z_2)|_{z_1 = z_2}\,\, + :U_1^s(z_2)V_2^s(z_2):\right) &\defnL
}$$
Explicitly, $L_{12}$ can be written in terms of the SYM fields as
\eqnn\explicitL
$$\eqalignno{
    L_{12}(z_1, z_2) & = \, :\Big[ - (k_2 \cdot A_1(z_1)) V_2(z_2) + W_1^{\alpha}(z_1) D_{\alpha} V_2(z_2) \cr
    & - {1 \over 4} (\lambda(z_2) \gamma_{mn} A_2(z_2)) {\cal F}_{1}^{mn}(z_1) \Big] e^{k_1 \cdot X(z_1) + k_2 \cdot X(z_2)}:.&\explicitL
}$$

Now, if we take the derivative with respect to $z_2$ and keep only divergent pieces, we find that
\eqnn\UpartialV
$$\eqalignno{
& U_1^s(z_1) \partial _2V^s_2(z_2) \sim \frac{k_1\cdot k_2}{z_{12}} (z_2 - z_1)^{-k_1 \cdot k_2} \Big(\frac{1}{z_{12}} L_{12}(z_2,z_2)+ \partial_1 L_{12}(z_1,z_2)|_{z_1=z_2}\,\, \cr
    & + :U^s_1(z_2)V^s_2(z_2):\Big) + (z_2 - z_1)^{-k_1 \cdot k_2} \left(\frac{1}{z_{12}^2} L_{12}(z_2,z_2)+ \frac{1}{z_{12}}\partial_2 L_{12}(z_2,z_2)\right) \cr
    & = (z_2 - z_1)^{-k_1 \cdot k_2} \Bigg(\frac{1 + s_{12}}{z_{12}^2} L_{12}(z_2,z_2) + \frac{1}{z_{12}} \partial_2 L_{12}(z_2,z_2)\,\, \cr
    & + \frac{s_{12}}{z_{12}} \left[ \partial_1 L_{12}(z_1,z_2)|_{z_1 = z_2}\,\,+ :U_1^s(z_2)V_2^s(z_2):\right].\Bigg)
}$$
We recall that $s_{12} = k_1 \cdot k_2$. We can perform the same computation for $\partial_1 V^s_1(z_1) U_2^s(z_2)$.
If we then match with the double pole obtained after applying $Q$ in \stringtheoryOPE,
we see that
\eqnn\contacttermstring
$$\eqalignno{
    Q D_{12}^{s} & = {1\over 2} (L_{12} + L_{21}), &\contacttermstring
}$$
which is precisely analogous to \ranktwoD. This leads us to split $L_{12}$
into its symmetric and antisymmetric pieces. Collecting now the single pole,
one finds that
\eqnn\QUranktwostring
$$\eqalignno{
    Q {U_{12}^{s} \over s_{12}} & = \partial \frac{1}{s_{12}}\left( L_{12} - L_{21} \right) + :U^s_{1}V^s_2: - :V^s_1 U^s_2: + \, Q(\dots). &\QUranktwostring
}$$
This is analogous to \ranktwoU\ up to the BRST
exact piece of the right-hand side, which is proportional to $Q D_{12}^s$ after using \contacttermstring. 
In other words, we can map $V_{12}$ to ${\left( L_{12} - L_{21} \right) \over s_{12}}$ 
and $U_{12} $ to $ {U_{12}^{s} \over s_{12}}$ up to contact terms. In particular, we see
that it is possible to construct the numerators of the particle vertex currents from the string theory
OPEs. Note that it is natural that
the equation for $U_{12}^{s}$ does not exactly reproduce \ranktwoU, since the definition of the
string OPE depends on what point we choose to perform the Laurent expansion, and the string OPE
also contains information about massive state excitations. Furthermore, we note that
if one produces a new set of gauge-equivalent vertices $U^{s \prime}$ and $V^{s \prime}$ such
that the corresponding $L_{ij}^{\prime}$ is antisymmetric, then the BRST cohomology of the single pole
of the $U^{s \prime}U^{s \prime}$ OPE exactly reproduces \ranktwoU. 

In the string amplitude computation, 
it can be shown that the BRST trivial piece $L_{(12)}$ cancels---roughly, 
the double pole piece $D^s_{12}$ can be incorporated into the $L_{ij}$ 
to update $L_{ij} \to L_{[ij]}$ \MafraNpoint. In this way, the string 
amplitude can be written solely in terms of $L_{[12]}$. This generalizes to more than two particles. 
The single poles of the string are (up to contact terms and BRST transformations) 
equal to the numerators of the Berends-Giele currents, while the higher-order poles are the contact terms. 
The interpretation of the worldline computation will be similar. Namely, the particle pinching operator $D_{ij}$ 
will act as a contact term in the amplitude that allows one to enhance the BRST properties of the kinematical 
numerators. As in string theory, the cancellation of these contact terms can be understood as coming from integration by parts identities.

\subsec Multiparticle worldline insertions

In this subsection, we will furnish expressions different expressions for the worldline insertions.
We start by giving the expression in terms of the SYM fields. First, we quote the 
results of \Mafraalgorithm\ for the SYM Berends-Giele currents in Lorenz gauge. 
These are given in terms of the multiparticle SYM fields \Mafraalgorithm, which are the operators written in usual roman letters. At rank one, ${\cal A}^{\alpha} = A^{\alpha}$, and similarly for other fields. At rank two,
\eqnn\SYMtwoparticle
$$\eqalignno{
{\cal A}_{\alpha}^{12} & = - {1 \over 2 s_{12}} \big[ A_{\alpha}^1 (k^1 \cdot A^2) + A^1_m (\gamma^m W^2)_{\alpha} - (1 2) \big], &\SYMtwoparticle \cr
{\cal A}_{m}^{12} & = -{1 \over 2 s_{12}} \Big[ A_{m}^{1} (k^1 \cdot A^2) + A_{n}^{1} F_{mn}^{2} - (W^{1} \gamma_m W^2) - (1  2) \Big], \cr
{\cal W}^{\alpha}_{12} & = -{1 \over 2 s_{12}} \Big[ W^{\alpha}_{1} (k_1 \cdot A_2) + W^{m \alpha}_{1} A_{m}^{2} + {1 \over 2} (\gamma^{rs} W^{1})^{\alpha} F_{rs}^2 - (1  2) \Big], \cr
{\cal F}^{mn}_{12} & = -{1 \over 2 s_{12}} \Big[ F^{mn}_{1} (k_1 \cdot A_2) + F^{p | mn}_{1} A_{p}^{2} + 2 F^{mp}_{1} F^{n}_{2p} + 4 \gamma^{[m}_{\alpha \beta} W_{1}^{n]\alpha} W_{2}^{\beta} - (1  2) \Big]. }
$$
Using these fields, we can construct the multiparticle worldline insertions,
\eqnn\Voneparticle
\eqnn\Vtwoparticle
$$\eqalignno{
V_1 & = \lambda^{\alpha} {\cal A}_{1\alpha}, & \Voneparticle \cr
V_{12} & = \lambda^{\alpha} {\cal A}_{12 \alpha}. &\Vtwoparticle
}
$$
We can also construct the worldline insertions and pinching operators. Comparing equations \U\ and \expansionU, we find
\eqnn\Uoneparticle
\eqnn\Utwoparticle
\eqnn\Dtwoparticle
$$\eqalignno{
U_1 & = p^m {\cal A}_{1m} + d_{\alpha} {\cal W}_1^{\alpha} + {1 \over 2} N^{mn} {\cal F}_{1mn}, & \Uoneparticle \cr
    U_{12} & = p^m {\cal A}_{12m} + d_{\alpha} {\cal W}_{12}^{\alpha} + {1 \over 2} N^{mn} {\cal F}_{12mn}, & \Utwoparticle \cr
    D_{12} & = {1 \over 2} ({\cal A}_1 \cdot {\cal A}_2 + {\cal W}_1 {\cal A}_2 + {\cal W}_2 {\cal A}_1). &\Dtwoparticle
}
$$
Note that the worldline insertions are proportional to inverse powers of the Mandelstam variables
and thus non-local. This procedure can be carried over to a higher number of external particles, 
and the BRST consistency conditions can be checked explicitly using the SYM equations of motion. 
We also note that $D_{12}$ has less propagators than the other objects at rank two---this again
generalizes to more particles.

The previous equations give one set of solution to the consistency conditions, but this solution
is not unique. Indeed, the worldline insertions are gauge-dependent objects, 
and it is useful to exploit this gauge freedom to introduce other representations of 
$V_P$ and $U_P$.
These will be constructed with the $b$-ghost operator introduced in section 2.1. 
First, we notice that the equations \allrankV\ mean that the unintegrated vertices $V_P$ are 
Berends-Giele currents \refs{\Mizera,\BG}. The natural ansatz for $V_P$ is
\eqnn\Vrecursion
$$\eqalignno{
V_{P} & = {1 \over s_P} \sum_{RQ=P} [U_R, V_Q], & \Vrecursion
}
$$
which determines the higher-rank $V_P$ in terms of lower point operators. 
We can check that this satisfies the correct consistency equation \allrankV. 
Acting with the BRST charge on \Vrecursion\ and using equation \allrankU, we find
\eqnn\QVrecursion
$$\eqalignno{
\{Q, V_{P}\} & = {1 \over s_P} \sum_{RQ=P} \Big\{ [[H,V_R], V_Q] + \sum_{R_1R_2=R} [U_{R_1} V_{R_2},V_Q] - [V_{R_1} U_{R_2}, V_Q] & \cr &  - \sum_{Q_1Q_2=Q}[U_R, V_{Q_1} V_{Q_2}] \Big\} & \cr 
&  = {1 \over s_P} \sum_{RQ=P} \Big\{- k_R \cdot k_Q V_R V_Q -\sum_{R_1R_2=R} V_{R_1}[U_{R_2},V_Q]  - \sum_{Q_1Q_2=Q}[U_R, V_{Q_1}] V_{Q_2} \Big\} & \cr 
& = - {1 \over s_P} \Big\{ \sum_{RQ=P} k_R \cdot k_Q V_R V_Q + \sum_{RQ_1Q_2=P} V_{R}[U_{Q_1},V_{Q_2}]  & \cr &  +\sum_{R_1R_2Q=P}[U_{R_1}, V_{R_2}] V_{Q} \Big\}. & \QVrecursion
}
$$
To obtain the second equality, we used $[H, V_R] = \partial_{\tau} V_R = (k_R \cdot P) V_R$, so that $[[H, V_R], V_Q] = - k_R \cdot k_Q V_R V_Q$, and also
applied the Jacobi identity for the remaining terms in the first and second line.
If we now plug in the iterative definition \Vrecursion\ into \QVrecursion, we see that
\eqnn\QVrecursionagain
$$\eqalignno{
\{Q, V_{P}\} & = - {1 \over s_P} \sum_{RQ=P} \Big\{ k_R \cdot k_Q + s_R + s_Q \Big\} V_R V_Q & \cr
& = - \sum_{RQ=P} V_R V_Q, & \QVrecursionagain
}
$$
where we remember that $s_P = k_P^2/2$. We have also used the identity $k_P^2 = (k_R + k_Q)^2 \Rightarrow s_P = k_R \cdot k_Q + s_R + s_Q $. This establishes that the recursive relation \Vrecursion\ satisfies the consistency condition \allrankV. 

Using the $b$-ghost, it is also possible to construct a recursion for the integrated vertices. 
For single particles, we define
\eqnn\Ubsingleparticle
$$\eqalignno{
\tilde{U}_i & = [b, V_i]. &\Ubsingleparticle
}
$$
We can use equation \btend\ to furnish an expression for $\tilde{U}_i$ in terms of the physical operators
\eqnn\Ubsingleparticleagain
$$\eqalignno{
\tilde{U}_i & = - \hat{{ \bf A}}_m(V_i) \Delta^m + \hat{\Delta}^m(V_i) { \bf A}_m . &\Ubsingleparticleagain
}
$$
It is possible to establish that $\tilde{U}_i$ is related to $U_i$ of equation \Utwoparticle\ 
by a BRST-exact piece \maxmaor \chandiavallilo. Further, one can easily check that such an object satisfies the 
BRST consistency \rankoneU\ by using the equation $\{Q, b\} = H$. Likewise, we define recursively
\eqnn\Vbrecursion
\eqnn\Ubrecursion
$$\eqalignno{
\tilde{V}_{P} & = {1 \over s_P} \sum_{RQ=P} [\tilde{U}_R, \tilde{V}_Q], & \Vbrecursion \cr
    \tilde{U}_{P} & = [b, \tilde{V}_{P}] = - \hat{{ \bf A}}_m(\tilde{V}_P) \Delta^m + \hat{\Delta}^m(\tilde{V}_P) { \bf A}_m.  &\Ubrecursion
}
$$
All of these fields satisfy the BRST consistency conditions necessary for the consistency of the 
coupled action. Using the tilded insertions, no pinching operators are necessary. Indeed, the equation
for the pinching operator say that its BRST variation is the symmetric piece of the multiparticle field,
\eqnn\pinchingbghost
$$\eqalignno{
    [Q, \tilde{D}_{P}] & = {1 \over 2} \sum_{RQ = P}([ \tilde{U}_R, \tilde{V}_Q] + [\tilde{U}_R, \tilde{V}_Q]) .  & \pinchingbghost
}$$
But for the tilded insertions, one can easily see that 
$[ \tilde{U}_R, \tilde{V}_Q] = - \hat{{ \bf A}}_m(\tilde{V}_R) \hat{\Delta}^m(\tilde{V}_Q) + \hat{\Delta}^m(\tilde{V}_R) \hat{{ \bf A}}_m(\tilde{V}_Q)$ 
is antisymmetric in $R\leftrightarrow Q$ (${\bf A}_m$ and $V_P$ are fermionic, while $\Delta_m$ is bosonic). Thus the RHS of the
last equation vanishes, and $[Q, \tilde{D}_{P}] = 0$. But this is defining equation on $\tilde{D}_{P}$---it 
is the only constraint we need to impose on the pinching operators to define 
a BRST invariant amplitude. So we can simply take $\tilde{D}_{P} = 0$ without losing any information. 

We can 
produce another yet set of insertions that are given in terms of the $b_0$ operator, which is the differential operator obtained by performing the substitution \momentumoperatormap\ to the $b$-ghost field \btend. We first go to Siegel gauge by imposing $b_0 (V^{\prime}_i) = 0$. If we now define $U^{\prime}_i = [b, V^{\prime}_i]$, we can check that
\eqnn\bzerovertices
$$\eqalignno{
    [U^{\prime}_i, V^{\prime}_j] & = b_0(V^{\prime}_i V^{\prime}_j).  & \bzerovertices
}$$
The same thing holds true if we recursively construct the higher rank vertices as
\eqnn\Vsiegelgauge
\eqnn\UVsiegelgauge
$$\eqalignno{
    V^{\prime}_P & = {1 \over s_P} \sum_{RQ = P} b_0(V^{\prime}_R V^{\prime}_Q), & \Vsiegelgauge \cr
    U^{\prime}_P & = [b, V^{\prime}_P]. & \UVsiegelgauge
}$$
Since $b_0(V^{\prime}_R V^{\prime}_Q) = [U^{\prime}_R, V^{\prime}_Q]$ in the Siegel gauge, it is clear that this set of multiparticle vertices satisfy the BRST consistency conditions. Moreover, since the $b_0$ operator is nilpotent, the unintegrated vertices satisfy
\eqnn\siegelgauge
$$\eqalignno{
    b_0(V^{\prime}_P) & = 0.  & \siegelgauge
}$$
for all $P$.

We can also study the pinching operators in the Siegel gauge. One finds
 \eqnn\Dsiegelgauge
$$\eqalignno{
    [Q, D_{P}^{\prime}] & = {1 \over 2} \sum_{RQ=P} b_0(V^{\prime}_R V^{\prime}_Q) + b_0(V^{\prime}_Q V^{\prime}_R) = 0,  & \Dsiegelgauge
}$$
which means that we can take $D_P^{\prime} = 0$. Here, we have used the anti-commutativity of the $V_P$ currents.

Using uniqueness in the BRST cohomology as in subsection (2.2), 
one can show that any two sets of worldline insertions satisfying the consistency 
conditions are related rank by rank by BRST-exact pieces and total derivatives. 
We will use this to show that the worldline amplitude prescription does not depend 
on which set of insertions one uses. The operators constructed with the $b$-ghost 
will prove particularly useful.

\subseclab \sectwoone

%------------------------------------------------------------------------------------------------------------------------------------------------------------------------------------------------ 4-POINT AMPLITUDE 10D--------------------------------------------------------------------------------------------------------------------------------------------------------------------------------------------------------------------------------------

\newsec Tree-level amplitude in 10D

 In this section, we will formulate a prescription to construct the tree-level scattering amplitudes of 10D ${\cal N}= 1$ SYM using the pure spinor superparticle, and then check that we obtain results that agree with the $\alpha^{\prime} \to 0$ limit of the open superstring. 

In SYM, tree-level amplitudes can be decomposed into color-ordered amplitudes. The $N$-point amplitude $A_N$ can be written as
\eqnn\colordec
$$\eqalignno{
&A_N(1, 2, \dots, N) = g^{N-2} \sum_{\sigma(1, \dots, N)/{\bZ_N}} \Tr(T^{\sigma_1} \dots T^{\sigma_N}) {\cal A}_N(\sigma_1, \sigma_2, \dots, \sigma_N),
}$$
where $g$ is the SYM coupling constant, the $T^a$ are a basis of the Lie-algebra, and $\sigma(1, \dots, N)$ are the non-cyclic permutations of the external indices $\{1, \dots, N\}$. The factor ${\cal A}_N(\sigma_1, \dots, \sigma_N)$ is the color-ordered amplitude, and only receives contributions from sub-diagrams in which the external legs are arranged in the order $\sigma_1, \dots, \sigma_N$.

In the worldline formulation, the correlator ${\cal A}_N(\sigma_1, \dots, \sigma_N)$ is written 
as a path integral over the fields defined on the particle worldline. The construction begins 
with the choice of the worldline, which can be the line connecting any pair of external states. 
These are interpreted as the initial and final states of the superparticle. 
The remaining legs are then represented by vertices ($V_P$ and $U_P$) coupling to the worldline, 
and can be seen as interactions of the superparticle with the SYM background. 
The amplitude is completely determined from the vertices and the worldline 
Green function, without the need to sew lower-point tree graphs to the worldline \Siegelworldline. 
We note however that the inserted vertices, being generally non-local, naturally
contain the information of lower-point tree graphs.

The two particles that define the worldline are represented by unintegrated vertex operators 
located at the endpoints of the worldline, $\tau = \pm\infty$. Due to the invariance of the worldline 
under global translations of $\tau$, one of the remaining insertions 
must also be represented by an unintegrated vertex. The amplitude does not depend on the position of
this third unintegrated insertion, and we fix it to be at $\tau = 0$. The remaining particles are 
represented by integrated vertices, whose proper time is integrated over the worldline. 
The amplitude ${\cal A}_N$ is given by the sum of all amplitudes with $N$ external states 
in the correct ordering. In particular, the interval of integration of the integrated vertices 
is such that this ordering is respected.

We will suppose that the external states are ordered clockwise as $1, 2, \dots, N$. When using this ordering, it is convenient to choose the worldline to be the line connecting $1$ and $N-1$, and also fix the position of the insertion $V_N$ (this is convenient because it ensures that we will not need to insert any multiparticle unintegrated vertices on the worldline). The propagation of the superparticle in the SYM background is described by the path integral 
\eqnn\path
$$\eqalignno{
{\cal A} & = \left\langle V_1(\infty) \exp\left\{\int d\tau\bU\right\} V_{N-1}(-\infty) \right\rangle & \cr
& = \sum {1 \over (N-2)!}\left\langle V_1(\infty) \left(\int d\tau\bU\right)^{N-3} V_N(0) V_{N-1}(-\infty) \right\rangle, & \path
}$$
where $\bU$ is the multiparticle superfield \U, described in the last section, and we have used the residual translation symmetry of the worldline to fix the $N$th insertion as an unintegrated vertex. Expanding $\bU$ into multiparticle states and restricting the sum to obey the correct number of external states and ordering, we find the color-ordered amplitude ${\cal A}_N$ as
\eqnn\Nptfunction
$$\eqalignno{
{\cal A}_N =\ & \sum_{|R_1| + \dots + |R_k| = N-3} \int d\tau_1 \dots d\tau_{k} & \Nptfunction\cr
&\times\langle V_1(\infty) (U_{R_1}+D_{R_1})(\tau_1) \dots (U_{R_k}+D_{R_k})(\tau_{k})V_N(0) V_{N-1}(-\infty)\rangle, &
}$$
\noindent where $U_P$ is the multiparticle integrated vertices and $D_R$ are the pinching operators, as defined in section 2. 

This correlator is evaluated by first contracting all momenta, using the worldline contractions \wick. This has the effect of promoting the canonical momenta to differential operators of the zero mode variables, up to the factors of $\sigma_{ij} = {1 \over 2}$sign$(\tau_i - \tau_j)$
\eqnn \momentumoperatormapagain
$$ \eqalignno{
 & p_m \to - \partial_m, \quad d_{\alpha} \to D_{\alpha}, \quad \omega_{\alpha} \to - \partial_{\lambda^{\alpha}}. & \momentumoperatormapagain
}$$
After that, one must perform the integration over the zero modes. One advantage of using the Hamiltonian action is that since there are no insertions of $\dot{x}^m$ on the worldline, one does not have to worry about possible ill-defined products of Dirac deltas.

\subsec Integration over zero modes

After getting rid of the canonical momenta, one still needs to integrate over $x^m$, $\theta^{\alpha}$ and $\lambda^{\alpha}$ to obtain a number out of equation \Nptfunction. The integration over $x^m$ furnishes a Koba-Nielsen-like measure on the space of $\tau_i$, as follows.

After contracting $p_m$, the amplitude \Nptfunction\ is proportional to the exponential factor
\eqnn\Exp
$$\eqalignno{
{\cal A}_N &\propto \int {\cal D} x \,  \exp\{- \int d\tau p_m \dot{x}^m \} \prod_{P} \exp\{ k_P \cdot x(\tau_P)\}. & \Exp
}
$$
This integral can be easily performed if we collect the plane wave factors under the integral, by writing $k_j \cdot x(\tau_j) = \int d\tau \delta(\tau - \tau_j) k_j \cdot x(\tau)$. The integral over the $x^m$ zero mode $x^m = $ const. furnish the momentum conservation delta function, which we omit. The integral over the non-zero modes furnishes, under integration by parts, a Dirac delta, as we can easily see above
\eqnn\deltap
$$\eqalignno{
&\int {\cal D} x \, \exp\left\{ \int d\tau \left( \dot{p}_m + \sum_j \delta(\tau - \tau_j) k_{jm} \right) x^m \right\} = \delta\left(\dot{p}_m + \sum_j \delta(\tau - \tau_j) k_{jm} \right).\ \ \ \ & \deltap
}
$$
We can use this Dirac delta to integrate over the $p_m$ field next. The equation
\eqnn\pdot
$$\eqalignno{
&\dot{p}_m(\tau) = - \sum_j \delta(\tau - \tau_j) k_{jm} & \pdot
}
$$
\noindent uniquely determines the canonical momenta $p_m$ once we impose that $p_m$ must be equal to the momentum of the boundary states at $\pm \infty$. We find
\eqnn\solp
$$\eqalignno{
& p^m(\tau) = - \sum_P \sigma_{\tau P} k_{P}^m. & \solp
}
$$
It can also be useful to employ the alternative expression
$$\eqalignno{
& p^m(\tau) = - {1\over2}\left(\sum_{\tau_P < \tau} k_{P}^m -  \sum_{\tau_P > \tau} k_{P}^m \right), & \solp
}
$$
which has the advantage of also being well-defined for $\tau = \tau_P$. We can use equation \solp\ to evaluate the factor of the path integral that comes from the kinetic energy of the particle. We obtain the following measure for the $\tau_P$ integration
\eqnn\KN
$$\eqalignno{
&\exp \left\{ {1\over2} \int d\tau p^2 \right\} =  \prod_i \exp\Bigg\{ {1\over2} (\tau_{i+1} - \tau_{i}) \Big( \sum_{j \leq i} k_{P_j} \Big)^2 \Bigg\}, & \KN
}
$$
which can also be written as
\eqnn\KNagain
$$\eqalignno{
&\exp \left\{ {1\over2} \int d\tau p^2 \right\} =  \prod_i \exp\Bigg\{ - {1\over2} \tau_i k_{P_i} \cdot \Big( \sum_{j <  i} k_{P_j} - \sum_{j >  i} k_{P_j} \Big) \Bigg\}. & \KNagain
}
$$
After integration against $\prod_i d\tau_i$, the factor \KN\ will furnish the channels of the amplitude corresponding to the worldline. Note that there are also channels corresponding to lower-point tree graphs sewed to the worldline; the corresponding Mandelstam variables come from the vertex operator insertions, as in equation \Exp.

To finish the computation of the amplitude, one must also perform the integration of the zero modes of $\theta^\alpha$ and $\lambda^\alpha$. This is carried out by projecting the remaining expression onto the pure spinor measure, which is the unique measure in the cohomology of the BRST charge with the correct ghost number of $+3$, given by
\eqnn\measure
$$\eqalignno{
&\langle (\theta \gamma^m \lambda) (\theta \gamma^n \lambda) (\theta \gamma^p \lambda) (\theta \gamma_{mnp} \theta) \rangle = 1. & \measure
}
$$

\subsec BRST invariance of the amplitude

One can prove that the amplitude \Nptfunction\ is BRST closed and invariant. 
This means that the value of the amplitude 
does not depend on which cohomology class the worldline insertions $U_P$ and $V_P$ belong to. 
Let us prove BRST closedness. First of all, for legibility, 
let us omit the $\tau$ dependence of the operators in equation \Nptfunction, and 
define the sum of the unintegrated vertex with the pinching operator as $\bU_P = U_P+D_P$. 
Then \Nptfunction\ can be rewritten as
\eqnn\Nptlegible
$$\eqalignno{
{\cal A}_N =\ & \sum_{|R_1| + \dots + |R_k| = N-3} \int d\tau_1 \dots d\tau_{k}\ \langle V_1\bU_{R_1}\bU_{R_2} \dots \bU_{R_k}V_NV_{N-1}\rangle.& \Nptlegible
}$$
Now let the BRST charge act on \Nptlegible. When $Q$ hits a unintegrated vertex $\bU_P$, using equations \allrankU\ and \allrankU\ one obtains
\eqnn\QbarU
$$\eqalignno{
& Q\bU_P = \partial V_P + \sum_{RQ=P}\left( U_{[R}V_{Q]} + [U_{\{R},V_{Q\}}] \right).&\QbarU
}$$
Thus, $Q {\cal A}_N$ is given by the following sum
\eqnn\QAN
$$\eqalignno{
Q{\cal A}_N & = -\sum_{k=1}^{N-3}\sum_{|R_1| + \dots + |R_k| = N-3}\int d\tau_1 \dots d\tau_{k}&\QAN\cr 
\times \sum_{i=1}^{k} \Bigg\langle & V_1 \bU_{P_1} \dots \bU_{P_{i-1}} \left(\partial V_{P_i} + \sum_{{R_i}{Q_i}={P_i}}\Big(U_{[{R_i}}V_{{Q_i}]} + [U_{\{{R_i}},V_{{Q_i}\}}]\Big)\right)\bU_{P_{i+1}}\cdots\bU_{P_k} V_{N}V_{N-1} \Bigg\rangle.
}$$

The integral over $\tau_i$ might be performed right away for the total derivative terms $\partial V_{P_i}$. Integrating over $\tau_i$, each of these terms in the last sum gives
\eqnn\intpartial
$$\eqalignno{
& - \sum_{k}\ \sum_i \int d\tau_1\cdots d\tau_k \langle V_1 \bU_{P_1}\cdots\bU_{P_{i-1}}\partial V_{P_i}\bU_{P_{i+1}}\cdots\bU_{P_k} V_{N}V_{N-1} \rangle & \intpartial\cr
= & - \sum_{k}\ \sum_i \int d\tau_1\cdots\hat{d\tau_i}\cdots d\tau_k \langle V_1 \bU_{P_1}\cdots\bU_{P_{i-1}}(\tau_{i-1}) V_{P_i}(\tau_{i-1}-\epsilon)\bU_{P_{i+1}}\cdots\bU_{P_k} V_{N}V_{N-1} \rangle &\cr
& + \int d\tau_1\cdots\hat{d\tau_i}\cdots d\tau_k \langle V_1 \bU_{P_1}\cdots\bU_{P_{i-1}} V_{P_i}(\tau_{i+1}+\epsilon)\bU_{P_{i+1}}(\tau_{i+1})\cdots\bU_{P_k} V_{N}V_{N-1} \rangle,
}$$
where $\hat{d\tau_i}$ indicates that the $d\tau_i$ differential is omitted. After relabeling $\tau_i$ and $\tau_{i+1}$, the last equation becomes
\eqnn\intpartialtwo
$$\eqalignno{
& - \sum_{k}\ \sum_i \int d\tau_1\cdots d\tau_{i-1} d \tau d\tau_{i+2} \dots d\tau_k \cr 
&  \times \langle V_1 \bU_{P_1}\cdots \left( V_{P_i}(\tau +\epsilon)U_{P_{i+1}}(\tau) - U_{P_i}(\tau)V_{P_{i+1}}(\tau-\epsilon) \right) \cdots\bU_{P_k} V_{N}V_{N-1} \rangle,&\intpartialtwo 
}$$
where we have dropped the terms in which $V_{P_1}$ approaches $+\infty$ and $V_{P_k}$ approaches $-\infty$. The subtraction $V_{P_i}(\tau+\epsilon)U_{P_{i+1}}(\tau) - U_{P_i}(\tau)V_{P_{i+1}}(\tau-\epsilon)$ can be written as an operator insertion on the worldline after appropriate normal ordering
\eqnn\normalord
$$\eqalignno{
& V_{P_i}(\tau+\epsilon)U_{P_{i+1}}(\tau) - U_{P_i}(\tau)V_{P_{i+1}}(\tau-\epsilon)&\cr
=\ & V_{P_i}U_{P_{i+1}}(\tau) - U_{P_i}V_{P_{i+1}}(\tau) + \half \Big( [U_{P_i},V_{P_{i+1}}] + [U_{P_{i+1}},V_{P_i}] \Big)(\tau) &\normalord\cr
=\ & V_{[P_i}U_{P_{i+1}]} + [U_{\{P_i},V_{P_{i+1}\}}].
}$$
Therefore, after summing equation \intpartial\ over all $P$ and all $i$ and normal ordering, we obtain
\eqnn\sumPandi
$$\eqalignno{
&\sum_{k=1}^{N-3}\sum_{|R_1| + \dots + |R_k| = N-3}\int d\tau_1 \dots d\tau_{k}&\sumPandi\cr
\times &\sum_i\sum_{R_iQ_i=P_i}\left\langle V_1\bU_{P_1}\cdots \bU_{P_{i-1}}\Big( U_{[R_i}V_{Q_i]} + [U_{\{R_i},V_{Q_i\}}]\Big)(\tau_i)\bU_{P_i+1}\cdots\bU_{P_k}V_NV_{N-1}\right\rangle,&
}$$
so the two kinds of terms in equation \QAN\ cancel, and the amplitude is indeed BRST closed. Pictorially, the BRST charge acts by cutting propagators on the amplitude, as in Figure 1. 

\ifig\QAfour{The action of $Q$ kills off internal propagators in the amplitude. 
There is a relative sign from whether the insertion is being collapsed to a vertex in its future or in its past. In this example, $2$ is being collapsed to its past in the first diagram and to its future in the second one.}
{\epsfxsize=.7\hsize\epsfbox{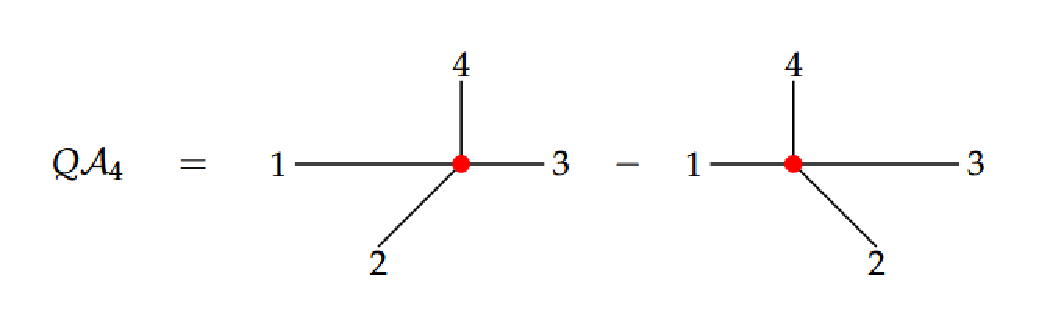}}

As mentioned before, the amplitude is also BRST invariant. This means that the value of the amplitude does not depend on what specific choice of vertex operators is made, as long as the choice is consistent in the sense of equations \allrankV\ and \allrankU. The demonstration follows the same steps as the proof performed above.

\subsec  4-point amplitude

In this subsection we will compute the 4-point amplitude to give a first example of the prescription detailed in the last subsections. The tree-level 4-pt amplitude contains only single-particle insertions and is given by
\eqnn\fouramplitude
$$\eqalignno{
{\cal A}_4 &= \int_{-\infty}^{\infty} d\tau_2 \langle V_1(\infty) U_2 (\tau_2) V_4(0) V_3(-\infty) \rangle. & \fouramplitude
}
$$
\noindent 
To evaluate this expression, we will first integrate over $x^m$, obtaining a kinematic measure such as in \KN. Then we will use the worldline contractions of the non-zero modes to promote the canonical momenta to differential operators as in \momentumoperatormapagain. Finally, we should project onto the pure spinor measure \measure\ to obtain the amplitude in terms of the external polarizations, but we leave this last step implicit. There exists software that automates this last step \MafraNpoint.

The $x^m$ dependence of the correlator in equation \fouramplitude\ comes from the plane wave factors of the worldline insertions. Using the results of subsection 3.1, the path integral over the $x^m$ and $p_m$ fields furnishes
\eqnn\KNfour
$$\eqalignno{
\int {\cal D} x {\cal D}p \,  & \exp\{- \int d\tau \left( p_m \dot{x}^m - {1 \over 2} p^2\right)\} \prod_{j = 1}^{4} \exp\{ k_j \cdot x(\tau_j)\} = \exp\{ (\sigma_{2i} s_{2i}) \tau_2\}, & \KNfour
}$$
where we have also defined $s_{ij} = k_i \cdot k_j$ and used the massless condition $k_i^2 = 0$. 
\noindent We can check that this factor gives the correct Mandelstam poles for the amplitude. The regions of integration $\tau_{2} > 0$ and $\tau_2 < 0$ correspond to the subdiagrams of Figure 2. 

\ifig\diagramfourpoint{The two subdiagrams that contribute to the 4-pt function. Note that time flows from the right to the left in our figures.}
{\epsfxsize=0.6\hsize\epsfbox{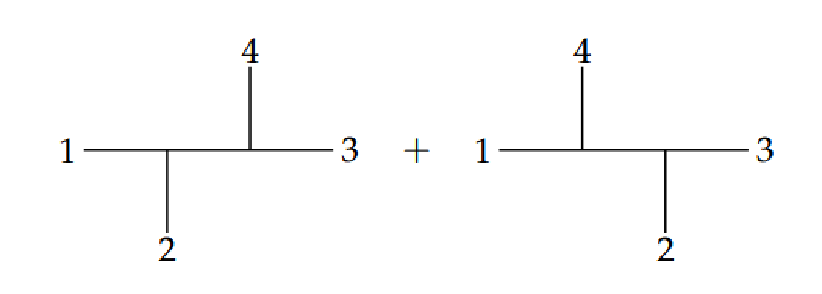}}

When integrated against $d\tau$, the kinematic measure gives the correct propagators corresponding to each of the diagrams in Figure 2. Indeed, we can notice that
\eqnn\mandelstams
\eqnn\mandelstamt
$$ \eqalignno{
\int_0^{\infty} d \tau_2 \exp\{ (\sigma_{2i} s_{2i}) \tau_2\} & = \int_0^{\infty} d \tau_2 \exp\{- s_{21} \tau_2\} = {1\over s_{21}},\ &\mandelstams \cr
\int_0^{\infty} d \tau_2 \exp\{ (\sigma_{2i} s_{2i}) \tau_2\} & = \int_{-\infty}^{0} d \tau_2 \exp\{+s_{23} \tau_2\} = {1\over s_{23}},&\mandelstamt
}
$$

\noindent
which is the correct result. This behavior generalizes to tree amplitudes with more external legs, but some of the poles come from the multiparticle vertices as in equation \expansionU. Now, we consider the contraction of the canonical momenta. Let us denote by $\hat{U}$ the differential operator obtained by substituting the canonical momenta of the integrated vertex $U$ using the rules of equations \momentumoperatormapagain. One then obtains for the 4-pt function \fouramplitude\ the following expression,
\eqnn\wickfour
$$\eqalignno{
{\cal A}_4 & =  \int d\tau_2 \exp\{ (\sigma_{2i} s_{2i}) \tau_2 \}  \left\langle \sigma_{21} \hat{U}_2(V_1) V_4 V_3 + \sigma_{23} V_1 \hat{U}_2(V_4) V_3 + \sigma_{24} V_1 V_4 \hat{U}_2(V_3) \right\rangle,\ \ \ \ \ \ &\wickfour
}
$$
where each vertex operator also carries a factor of $\exp\{ k \cdot x\}$ where $x^m$ is the constant mode of the position variable. For concreteness, we also perform the $\tau_2$ integration, obtaining
\eqnn\fourpointwithU
$$\eqalignno{
{\cal A}_4 & = {1 \over 2 s_{21}} \langle- \hat{U}_2(V_1) V_4 V_3 + V_1 \hat{U}_2(V_4) V_3 + V_1 V_4 \hat{U}_2(V_3) \rangle \cr
&+ {1 \over 2 s_{23}} \langle- \hat{U}_2(V_1) V_4 V_3 - V_1 \hat{U}_2(V_4) V_3 + V_1 V_4 \hat{U}_2(V_3) \rangle. &\fourpointwithU
}
$$
Each of the terms in the last equation corresponds to one of the subdiagrams in Figure 2. It would be natural to expect that the summand corresponding to the channel $s_{21}$ would have a numerator where $\hat{U}_2$ acts only on $V_1$ (as Figure 2 suggests), and similarly for the channel $s_{23}$. In our partial result, however, this is not true, unless it is possible to somehow integrate by parts the action of $\hat{U}_2$ in equation \fourpointwithU. That this is possible in the pure spinor formalism is not clear at first, because the integration over $\theta^{\alpha}$ and $\lambda^{\alpha}$ is obtained by projection onto the pure spinor measure \measure\ instead of ordinary integration. In Appendix A we will argue that this is indeed allowed.

After performing such integration by parts, we rewrite the 4-pt amplitude as
\eqnn\fourpointwithUagain
$$\eqalignno{
{\cal A}_4 & = - {1 \over s_{21}} \langle \hat{U}_2(V_1) V_4 V_3 \rangle + {1 \over s_{23}} \langle V_1 V_4 \hat{U}_2(V_3) \rangle. &\fourpointwithUagain
}
$$
It is of interest to understand the BRST cohomology of the numerators appearing in tree amplitudes. The last equation can be written as
\eqnn\fourpointBRSTcohomology
$$\eqalignno{
{\cal A}_4 & = {1 \over s} n_s + {1 \over u}  n_u,&\fourpointBRSTcohomology
}
$$
where we introduced the kinematical numerators
\eqnn\kin
$$\eqalignno{
& \hat{U}_2(V_1) V_4 V_3= - n_{s},\quad V_1 \hat{U}_2(V_4) V_3 = n_t, \quad V_1 V_4 \hat{U}_2(V_3) = n_u. & \kin
}$$
The requirement that the amplitude be BRST closed fixes the BRST cohomology of the kinematical numerators $n_i$. Indeed, the 4-pt amplitude can only obey $Q{\cal A}_4 = 0$ for every value of the Mandelstam variables if there exists some operator ${\cal O}$ such that
\eqnn\numeratorsBRSTcohomology
$$\eqalignno{
Qn_s & = - s {\cal O}, \quad Qn_u = u {\cal O}. &\numeratorsBRSTcohomology
}
$$
The objects satisfying equation \numeratorsBRSTcohomology\ were studied thoroughly in \MafraNpoint. These are the BRST building blocks, that we review in Appendix B. We use the BRST building blocks to organize our expressions.

We now return to equation \fourpointwithUagain\ and compute the action of the integrated vertex on the unintegrated vertex, $\hat U_i(V_j)$. Using the equations of motion of SYM, one finds
\eqnn\UV
$$\eqalignno{
&\hat{U}_i(V_j) = - \Big( (A_i \cdot k_j) V_j + (W_i \gamma^m \lambda) A_{jm} + Q( W_i A_j) \Big). &\UV
}$$
\noindent This can be further rewritten if we use the identity
\eqnn\QD
$$\eqalignno{
Q( A_1 \cdot A_2 ) & = k_1 \cdot A_2 V_1 + (\lambda \gamma^m W_1) A_2^m + (1\leftrightarrow2). &\QD
}$$
which can be checked explicitly from \Utwoparticle. We find 
\eqnn\UVagain
$$\eqalignno{
&{1\over s_{12}}\hat{U}_2(V_1) = \lambda^{\alpha} A_{\alpha12} + {1\over s_{12}} QD_{12} = V_{21} + {1\over s_{12}} Q D_{12}. &\UVagain
}$$
Plugging this into \fourpointwithUagain, the BRST-exact piece decouples trivially after integration by parts. So the four-point function can be written as
\eqnn\finalfourpt
$$\eqalignno{
{\cal A}_4 & = - \langle V_{12} V_3 V_4 \rangle - \langle V_1 V_{23} V_{4}\rangle, &\finalfourpt
}$$
where we also used the fact that $V_{ij}$ is antisymmetric in $i$ and $j$ (notice also that $V$ is fermionic). We have thus seen that the BRST exact pieces decouple and the amplitude can be written in terms of the Berends-Giele currents $V_P$, which are constructed with the BRST building blocks (see appendix B). This is the correct result obtained by the $\alpha^{\prime} \to 0$ limit of the superstring \MafraNpoint.

This simple argument for the decoupling of the BRST exact piece of $\hat{U}(V)$ is no longer true at higher points, because there the BRST charge can hit other integrated vertices after integration by parts. This will produce residual terms, but they cancel exactly with the contributions coming from the pinching operators in equation \Nptfunction.

%------------------------------------------------------------------------------------------------------------------------------------------------------------------------------------------------------- 5-POINT AMPLITUDE 10D------------------------------------------------------------------------------------------------------------------------------------------------------------------------------------

\subsec  5-point amplitude

For the color ordering $12345$, the tree level 5-point amplitude coming from \Nptfunction\ is given by
\eqnn\fivepoint
$$\eqalignno{
{\cal A}_5  & = \int_{-\infty}^{\infty} d\tau_2 \int^{\tau_2}_{-\infty} d\tau_3 \Big\langle V_1(\infty) U_2(\tau_2) U_3(\tau_3) V_5(0) V_4(-\infty) \Big\rangle & \cr 
& \ \ \ \ \ + \int_{-\infty}^{\infty} d\tau \langle V_1(\infty) \left( U_{23}(\tau) + D_{23}(\tau) \right) V_5(0) V_4(-\infty)\Big\rangle, & \fivepoint
}$$
\noindent where in the second line we included the contribution from the two-particle integrated vertex $U_{23}$ of equation \Utwoparticle\ and also from the pinching operator $D_{23}$ of equation \Dtwoparticle. 

As before, the $x^m$ path integral gives a kinematical measure that is integrated in $\tau$ to give the propagators of the amplitude. After this is done, one integrates over the canonical momenta, which has the effect of promoting the integrated vertices to differential operators. Computing the correlator, we find a sum of sixteen contractions in the first line of equation \fivepoint\ and three in the second line. We can simplify this expression by integrating the action of the integrated vertices by parts, as before. To do this, we must first perform a gauge transformation of the vertex operators that takes advantage of the properties of the $b$-ghost. 

The gauge transformation that we will perform is
\eqnn\vertextransformation
$$\eqalignno{
& U_{P} \mapsto \tilde{U}_P, & \vertextransformation
}$$
where the tilded vertices, defined in subsection 2.5, are constructed from the $b$-ghost operator. In particular, the pinching operator turns out to be BRST closed and can be taken to be zero without loss of generality
as explained below equation \pinchingbghost. After this gauge transformation is performed, the amplitude can be written as
\eqnn\fivepointtilde
$$\eqalignno{
{\cal A}_5  & = \int_{-\infty}^{\infty} d\tau_2 \int^{\tau_2}_{-\infty} d\tau_3 \Big\langle V_1(\infty) \tilde U_2(\tau_2) \tilde U_3(\tau_3) V_5(0) V_4(-\infty) \Big\rangle & \cr 
& \ \ \ \ \ + \int_{-\infty}^{\infty} d\tau \langle V_1(\infty) \tilde U_{23}(\tau) V_5(0) V_4(-\infty)\Big\rangle.  &\fivepointtilde
}$$
The first term in the equation above corresponds to the three diagrams in the first line of Figure 3, while the second term corresponds to the diagrams in the second line. 

\ifig\diagramfivepoint{The diagrams that contribute to the 5-pt function. We have split the diagrams according to the region of integration. On the first line, all the diagrams have only single particle insertions, while in the second line the insertions of $U_{23}$ are represented by a subtree. Finally, the third line includes contact terms, represented by the pinching operators $D_{23}$.}
{\epsfxsize=0.7\hsize\epsfbox{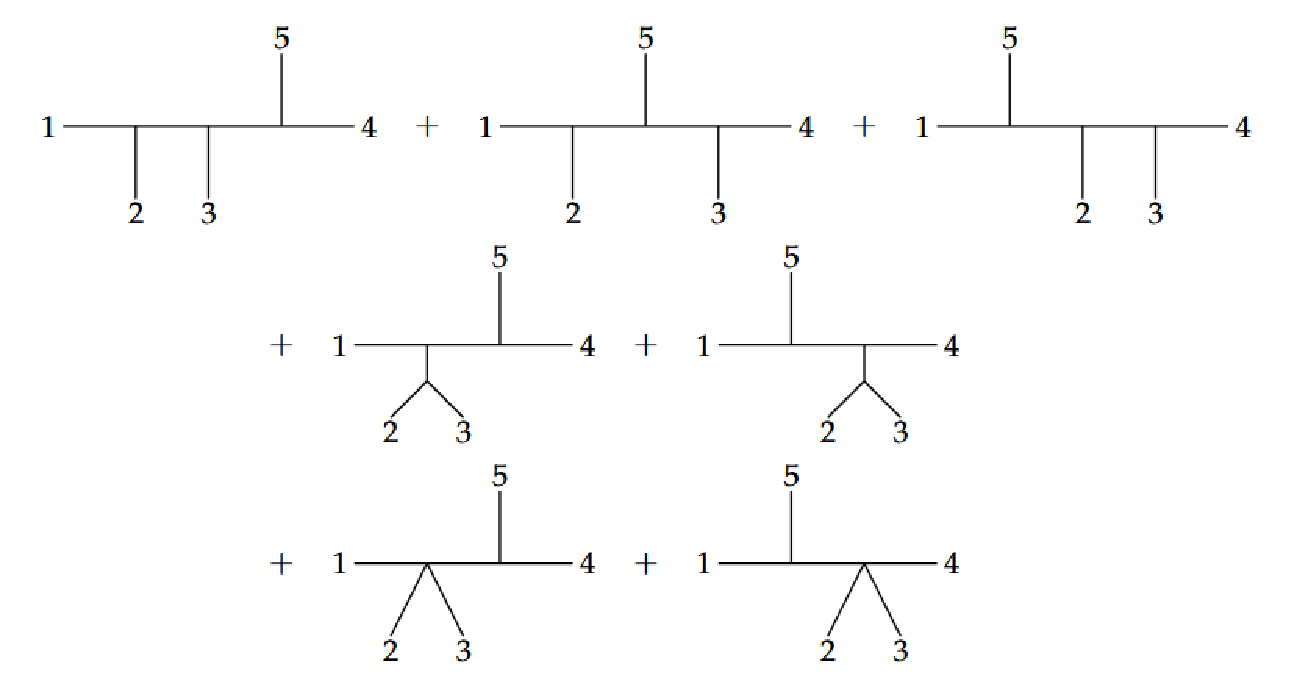}}

We shall now proceed to the computation of the correlator. The path integral over the canonical momenta is equivalent to 
replacing the canonical momenta in the integrated vertex insertions with the corresponding differential operators, up to $\sigma_{ij}$ factors. For instance
\eqnn\replacingP
$$\eqalignno{
& \langle p_m(\tau_1){\cal O}(x(\tau_2)) \rangle = \sigma_{12}\langle \partial_m{\cal O}(x(\tau_2)) \rangle. &\replacingP
}$$
Performing this substitution in equation \fivepointtilde, one defines the hatted operators $\hat{\tilde U}_i$ and $\hat{\tilde U}_{ij}$. Using the properties of the $b$ ghost, it is possible to integrate these operators by parts (see Appendix A), which greatly simplifies the computation of the correlator. To illustrate, let us take the first integral in Equation \fivepointtilde\ and consider the interval of integration $\infty > \tau_2 > \tau_3 > 0$, corresponding to the first diagram of Figure 3. The corresponding piece of the amplitude is equal to
\eqnn\fivepointfirstdiagram
$$\eqalignno{
&\int_{0}^{\infty} d\tau_2 \int_{0}^{\tau_2} d\tau_3 e^{-\tau_2s_{12}}e^{-\tau_3(s_{13}+s_{23})} \Big\langle V_1(\infty) \tilde U_2(\tau_2) \tilde U_3(\tau_3) V_5(0) V_4(-\infty) \Big\rangle & \cr
= & -\half\int_{0}^{\infty} d\tau_2 \int_{0}^{\tau_2} d\tau_3 e^{-\tau_2s_{12}}e^{-\tau_3(s_{13}+s_{23})} \Big\langle \big(\hat{\tilde U}_2(V_1)\tilde U_3 V_5V_4 - V_1\hat{\tilde U}_2(\tilde U_3 V_5V_4) \big)\Big\rangle &\cr
= & - \int_{0}^{\infty} d\tau_2 \int_{0}^{\tau_2} d\tau_3 e^{-\tau_2s_{12}}e^{-\tau_3(s_{13}+s_{23})} \Big\langle \hat{\tilde U}_2(V_1)\tilde U_3 V_5V_4 \Big\rangle, &\fivepointfirstdiagram
}$$
where we integrated the action of $\hat{\tilde U}_2$ by parts in the last line. Using the same argument for the action of $\tilde U_3$ and performing the integral over the kinematic measure, one obtains the value of the correlator together with the corresponding kinematical poles of the amplitude
\eqnn\firstpole
$$\eqalignno{
&{1\over s_{12}s_{45}}\langle \hat{\tilde U}_3(\hat{\tilde U}_2(V_1))V_5V_4 \rangle. &\firstpole
}$$
This computation is conceptually important because it demonstrates how the contractions coming from the worldline can be arranged as Berends-Giele currents. A useful pictorial interpretation of this result is that, in the corresponding diagram of this correlator $-$ specifically the first diagram of figure 3 $-$ each integrated vertex acts only on the leg where it sits.

If we repeat this procedure to all channels, we obtain the following expression for the amplitude
\eqnn\prefivepoint
$$\eqalignno{
{\cal A}_5 = \ & {1\over s_{12}s_{45}}\langle \hat{\tilde U}_3(\hat{\tilde U}_2(V_1))V_5V_4 \rangle + {1\over s_{12}s_{34}}\langle \hat{\tilde U}_2(V_1)V_5\hat{\tilde U}_3(V_4) \rangle + {1\over s_{15}s_{34}}\langle V_1V_5 \hat{\tilde U}_3(\hat{\tilde U}_2(V_4)) \rangle &\cr
&+ {1\over s_{45}}\langle \hat{\tilde U}_{23}(V_1)V_5V_4 \rangle + {1\over s_{15}}\langle V_1V_5\hat{\tilde U}_{23}(\hat{\tilde U}_4) \rangle. &\prefivepoint
}$$
To compare with the previous result for SYM amplitudes in the pure spinor formalism of \MafraNpoint, we will write this result in terms of the BRST building blocks $T_{ij}$ and $T_{ijk}$. As it turns out, any set of composite operators satisfying the right BRST cohomology can be used to express the amplitude, and different composite operators are related by BRST-exact pieces and contact terms that cancel in the full color-ordered amplitude. The BRST cohomology and symmetries of $T_{ij}$ and $T_{ijk}$ are enough to obtain our result.

The BRST building blocks are composite fields that satisfy the following BRST cohomology \MafraNpoint
\eqnn\QTij
\eqnn\QTijk
$$\eqalignno{
&\{Q,T_{12}\} = s_{12}V_1V_2, &\QTij\cr
&\{Q,T_{123}\} = s_{123}T_{12}V_3 + s_{12}(T_{13}V_2 - T_{12}V_3 + V_1T_{23}), &\QTijk
}$$
and the symmetries $T_{(12)} = T_{(12)3} = T_{123} + cyclic = 0$. In order to express our result \prefivepoint\ in terms of the BRST building blocks, we note that the composite vertices $\hat{\tilde U}_i(V_j)$, $\hat{\tilde U}_i(\hat{\tilde U}_j(V_k))$ and $\hat{\tilde U}_{ij}(V_k)$ satisfy the same relation under BRST transformation, namely
\eqnn\QUij
\eqnn\QUijk
$$\eqalignno{
& \{Q,\hat{\tilde U}_2(V_1)\} = s_{12}V_2V_1, &\QUij\cr
& \{Q,\hat{\tilde U}_3(\hat{\tilde U}_2(V_1))\} = s_{123}\hat{\tilde U}_2(V_1)V_3 + s_{12}(\hat{\tilde U}_3(V_2)V_1 + V_2\hat{\tilde U}_3(V_1) -V_3\hat{\tilde U}_2(V_1)). &\QUijk
}$$
This set of equations means that the composite vertices can be mapped to the BRST building blocks up to contact terms. Let us see how this plays out. At two-particles, one can see that $\{Q,\hat{\tilde U}_2(V_1) - T_{21}\}=0$. The uniqueness of the BRST cohomology then implies that the composite vertex is related to this building block at this rank by a BRST exact piece; i.e. there exists $E_{21}$ such that
\eqnn\ranktwoUandT
$$\eqalignno{
& \hat{\tilde U}_2(V_1) = T_{21} + [Q,E_{21}].&\ranktwoUandT
}$$
The anti-symmetry of both the composite vertex and the BRST building block implies that $E_{ij}$ is also antisymmetric, $E_{(ij)}=0$. 
For the three-particle composite fields, one can find from \QTijk, \QUijk\ and \ranktwoUandT,
\eqnn\rankthreeQUandQT
$$\eqalignno{
& \{Q,\hat{\tilde U}_3(\hat{\tilde U_2}(V_1))\} = \{Q, T_{123} -s_{123}V_3E_{21} + s_{21} (E_{32}V_1 - E_{31}V_2 + E_{21}V_3)\}. &\rankthreeQUandQT
}$$
The uniqueness of the BRST cohomology then implies that
\eqnn\rankthreeUandT
$$\eqalignno{
& \hat{\tilde U}_3(\hat{\tilde U_2}(V_1)) = T_{123} -s_{123}V_3E_{21} + s_{21} (E_{32}V_1 - E_{31}V_2 + E_{21}V_3) + [Q,\cdots]. &\rankthreeUandT
}$$
The terms that come multiplied by Mandelstam variables are named contact terms. The result thus obtained is that the composite field is equal to the BRST building blocks plus contact terms and BRST-exact pieces. The $[Q_0, \cdots]$ in the
last equation does not contribute to the 5-pt function, but if we were interested in the higher point functions, we would need to include them, and they would give contact terms for the higher-order composite fields. Proceeding similarly, one finds that
\eqnn\UcontactT
$$\eqalignno{
& s_{23}\hat{\tilde U}_{23}(V_1) = T_{321} - s_{123}V_1E_{23} + s_{23} (E_{12}V_3 - E_{13}V_2 + E_{23}V_1) + [Q,\cdots]. &\UcontactT
}$$

To write the result \prefivepoint\ in terms of the BRST building blocks, we plug the equations \ranktwoUandT, \rankthreeUandT\ and \UcontactT. The contact terms $E_{ij}$ cancel between diagrams and the final result for the 5-point function is
\eqnn\fivepointfinal
$$\eqalignno{
{\cal A}_5 = \ & {\langle T_{123}V_5V_4\rangle \over s_{12}s_{45}} + {\langle T_{12}T_{34}V_5\rangle \over s_{12}s_{34}} + {\langle V_1T_{432}V_5\rangle\over s_{15}s_{34}} + {\langle T_{321}V_4V_5\rangle\over s_{23}s_{45}} + {\langle V_1T_{234}V_5\rangle\over s_{23}s_{15}}. &\fivepointfinal
}$$

This result agrees with the low energy limit of the pure spinor superstring disk amplitude \MafraNpoint. The same result can be obtained without the $b$-ghost by writing down all the contractions coming from \fivepoint, and using the BRST cohomology to obtain an expression similar to \prefivepoint, but including pinching operators. Then one can again argue that the pinching operators cancel with the BRST exact piece of the multiparticle vertices, and the result is again that of eq. \fivepointfinal.

%---------------------------------------------------------------------------------------------------------------------------------------------------------------------------------------------------------------------------------------------------------------------------------------------------------------------------------------------- N-POINT AMPLITUDE 10D------------------------------------------------------------------------------------------------------------------------------------------------------------------------------------------------------------------------------------------------------------------------------------------------------------------------------------------------------

\subsec  N-point amplitude

The $N$-point SYM tree-level amplitude was computed in \maframultiparticles\ using the Berends-Giele currents constructed from the BRST building blocks. The result is given by
\eqnn\NpointMafra
$$\eqalignno{
{\cal A}_N^{\prime} = & \sum_{j=1}^{N-2} \langle M_{1 \dots j} M_{j+1 \dots N-1} V_N \rangle, & \NpointMafra
}$$
where the Berends-Giele currents $M_{i \dots j}$ are defined iteratively through their BRST cohomology. They satisfy
\eqnn\MoperatorBRSTcohomology
$$\eqalignno{
Q M_{1 \dots p} & = \sum_{j=1}^{p-1} M_{1 \dots j} M_{j+1 \dots p}, & \MoperatorBRSTcohomology
}$$
with $M_i = V_i$. One should notice that this is the same recursion relation that defines the multiparticle unintegrated vertices as in equation \allrankV. We will now show that the worldline amplitude of equation \Nptfunction\ can be written in the same way as in equation \NpointMafra, using a potentially different set of Berends-Giele currents. We will then use the uniqueness in the BRST cohomology to establish that the value of the amplitude is independent of the representation of the Berends-Giele currents, proving that the worldline prescription gives the same result as equation \NpointMafra. For convenience, we will write down the worldline amplitude prescription (equation \Nptfunction) again
\eqnn\Nptfunctionagain
$$\eqalignno{
{\cal A}_N =\ & \sum_{|R_1| + \dots + |R_k| = N-3} \int_{\tau_j > \tau_{j+1}} d\tau_{1} \dots d\tau_{k} & \cr
&\times\langle V_1(\infty) V_N(0) V_{N-1}(-\infty) (U_{R_1}+D_{R_1})(\tau_1) \dots (U_{R_k}+D_{R_k})(\tau_{k})\rangle. & \Nptfunctionagain
}$$
Gauge invariance allows us to rewrite ${\cal A}_N$ in terms of the $b$-ghost vertex insertions, which are denoted by a tilde, as follows
\eqnn\Nptbghost
$$\eqalignno{
    {\cal A}_N = &\sum_{k=1}^{N-3}\sum_{\sum_{i=1}^k |P_i| = N-3} \int_{\tau_j > \tau_{j+1}} d\tau_{1} \dots d\tau_{k} \langle V_1 \tilde{U}_{P_1} \dots \tilde{U}_{P_k} V_N V_{N-1} \rangle. & \Nptbghost
}
$$
The worldline wick contractions promote the vertex insertions to differential operators as explained in subsection 3.4. Due to the properties of the $b$-ghost, we can integrate the action of these operators by parts, obtaining a sum of strings of operators acting on $V_1$ or $V_{N-1}$ according to where $\tau_i$ is greater or smaller than zero
\eqnn\Nptbghosttwo
$$\eqalignno{
    {\cal A}_N & = \sum_{k=1}^{N-3}\sum_{\sum_{i=1}^k |P_i| = N-3} \sum_{j} (-1)^j\int_{\tau_j > 0 > \tau_{j+1}} d\tau_{1} \dots d\tau_{k} \cr 
    & \times \left\langle \left( \hat{\tilde{U}}_{P_1} \dots \hat{\tilde{U}}_{P_j}  V_1 \right) V_N \left(\hat{\tilde{U}}_{P_{j+1}} \dots \hat{\tilde{U}}_{P_k} V_{N-1} \right) \right\rangle. & \Nptbghosttwo
}
$$
Although this expression seems complicated, we can readily identify it as a sum of Berends-Giele currents as defined in \MoperatorBRSTcohomology. For instance, summing over all partitions of $j+1, \dots, N-2$, we find
\eqnn\berendsgielefromamplitude
$$\eqalignno{
    \sum_{\sum P_i = j+1, \dots, N-3}\int_{0 > \tau_{j+1}} d\tau_{j+1} \dots d\tau_{k} \left(\hat{\tilde{U}}_{P_{j+1}} \dots \hat{\tilde{U}}_{P_k} V_{N-1} \right)    & = \tilde V_{j+1 \dots N-1},& \berendsgielefromamplitude
}
$$
where the multiparticle field $\tilde V_{j+1 \dots N-1}$ is defined as in \Vrecursion\ $-$ the factors of $1/s_{P}$ come from the proper time integration. Thus equation \Nptbghosttwo\ can be written as
\eqnn\Nptbghostthree
$$\eqalignno{
    {\cal A}_N & = - \sum_{j=1}^{N-2} \langle \tilde{V}_{1 \dots j} \tilde{V}_{j+1 \dots N-1} V_N \rangle, & \Nptbghostthree
}
$$
where we used the shuffle symmetry of the Berends-Giele currents \Mizera. A similar argument can be constructed without using the $b$-ghost, but one needs to subtract contact terms appropriately when defining the Berends-Giele currents to guarantee the correct BRST cohomology. 

We will now prove that every set of Berends-Giele currents gives rise to the same scattering amplitude. Suppose that
\eqnn\amplitudecurrents
$$\eqalignno{
    {\cal A}_N & = - \sum_{j=1}^{N-2} \langle V_{1 \dots j} V_{j+1 \dots N-1} V_N \rangle, & \amplitudecurrents
}$$
for some set of Berrends-Giele currents satisfying
\eqnn\berendsBRST
$$\eqalignno{
    Q V_P & = - \sum_{RQ=P} V_R V_Q. & \berendsBRST
}$$
Suppose further that there exists another set of Berrends-Giele currents and construct ${ \cal A}^{\prime}_N$ as
\eqnn\amplitudecurrentstwo
$$\eqalignno{
    {\cal A}^{\prime}_N & = - \sum_{j=1}^{N-2} \langle V^{\prime}_{1 \dots j} V^{\prime}_{j+1 \dots N-1} V_N \rangle, & \amplitudecurrentstwo
}$$
Suppose further that $V^{\prime}_i = V_i$. The two sets of Berrends-Giele currents will be related to each other by contact terms and BRST-exact pieces. For instance
\eqnn\berendsgieletwo
$$\eqalignno{
    Q(V^{\prime}_{12} - V_{12}) & = 0 \Rightarrow V^{\prime}_{12} = V_{12} + Q E_{12}, & \berendsgieletwo
}
$$
for some $E_{12}$, since the BRST cohomology is trivial for off-shell fields by what
is argued in subsection (2.2). This implies that
\eqnn\berendsgielethree
$$\eqalignno{
    Q V^{\prime}_{123} & = Q V_{123} + Q( - E_{12} V_3 + V_1 E_{23}) \cr
    \Rightarrow V^{\prime}_{123} & = V_{123} - E_{12}V_3 + V_{1}E_{23} + Q(E_{123}), & \berendsgielethree
}$$
for some $E_{123}$. This process can be continued for all ranks of the currents. Schematically
\eqnn\berendsgieleP
$$\eqalignno{
    V^{\prime}_P & = V_P + Q(E_P) + contact\ terms. & \berendsgieleP
}$$
Now consider ${\cal A}^{\prime}_N$, and expand the $V^{\prime}_P$ appearing in the expansion as $V_P$, plus contact terms and BRST exact pieces. We will fix
$P = 1\dots k$ and track the term proportional to $E_P$ to see that it cancels. This term comes from $V^{\prime}_P$ and also from every higher rank current $V^{\prime}_{P Q}$, where $Q = (k+1) \dots j$ for some $j$. 
Let us study the case $Q = k+1$,
\eqnn\berendsgielePtwo
$$\eqalignno{
    Q V^{\prime}_{P(k+1)} & = - V^{\prime}_{P} V^{\prime}_{k+1} + \dots \cr
    & = - V_P V^{\prime}_{k+1} - Q E_{P} V^{\prime}_{k+1} + \dots \cr
    & = - V_P V^{\prime}_{k+1} - Q( E_{P} V^{\prime}_{k+1}) + \dots, \cr
    \Rightarrow V^{\prime}_{P(k+1)} & = V_{P(k+1)} - E_P V^{\prime}_{k+1} + \dots, & \berendsgielePtwo
}$$
where the $\dots$ include all terms that do not have $E_P$. Similarly
\eqnn\berendsgielePthree
$$\eqalignno{
    Q V^{\prime}_{P(k+1)(k+2)} & = - V_{P} V^{\prime}_{(k+1)(k+2)} - Q(E_P V^{\prime}_{(k+1)(k+2)}) + \dots \cr
    \Rightarrow V^{\prime}_{P(k+1)(k+2)} & = V_{P(k+1)(k+2)} - E_P V^{\prime}_{(k+1)(k+2)} + \dots. & \berendsgielePthree
}$$
In general, one finds (for $Q = (k+1) \dots j$)
\eqnn\berendsgielePfour
$$\eqalignno{
    Q V^{\prime}_{P Q} & = - V_P V^{\prime}_Q - Q(E_P V^{\prime}_Q) + \dots \cr
    \Rightarrow V^{\prime}_{PQ} & = V_{PQ} - E_{P} V^{\prime}_Q + \dots. & \berendsgielePfour
}$$
Using this equation, the terms proportional to $E_P$ are easily seen to cancel in ${\cal A}^{\prime}_N$.
Indeed, fix $P = 1\dots k$. Then (here, $Q$ is a shortcut for $((k+1)\dots N-1)$, and $Q_1 Q_2$ run over the deconcatenation sum of $Q$)
\eqnn\proofNpoint
$$\eqalignno{
    {\cal A}^{\prime}_N & = - \sum_{j=1}^{N-2} \langle V^{\prime}_{1 \dots j} V^{\prime}_{j+1 \dots N-1} V_N \rangle \cr
    & = -  \langle V^{\prime}_{P} V^{\prime}_{Q} V_N \rangle - \sum_{Q_1 Q_2 = Q} \langle V^{\prime}_{PQ_1} V^{\prime}_{Q_2} V_N \rangle + \dots \cr
    & = - \langle V_{P} V^{\prime}_Q V_N \rangle - \langle Q E_P V^{\prime}_Q V_N \rangle - \sum_{Q_1 Q_2=Q} \langle V^{\prime}_{PQ_1} V^{\prime}_{Q_2} V_N \rangle + \dots \cr
    & = - \langle V_{P} V^{\prime}_Q V_N \rangle + \langle E_P QV^{\prime}_Q V_N \rangle - \sum_{Q_1 Q_2=Q} \langle V^{\prime}_{PQ_1} V^{\prime}_{Q_2} V_N \rangle + \dots \cr
    & = - \langle V_{P} V^{\prime}_Q V_N \rangle -  \sum_{Q_1 Q_2=Q} \langle E_P V^{\prime}_{Q_1} V^{\prime}_{Q_2} V_N \rangle - \sum_{Q_1 Q_2=Q} \langle V^{\prime}_{PQ_1} V^{\prime}_{Q_2} V_N \rangle + \dots \cr
    & = - \langle V_{P} V^{\prime}_Q V_N \rangle - \sum_{Q_1 Q_2=Q} \langle V_{PQ_1} V^{\prime}_{Q_2} V_N \rangle + \dots, & \proofNpoint
}$$
where the $\dots$ contain no term proportional to $E_P$. A similar argument can be constructed for the cancellation of the $E_P$ where $P \neq 1\dots k$ (for instance by using shuffle symmetry). After all of the $E_P$ terms are canceled, one finds that
\eqnn\proofNpointagain
$$\eqalignno{
    { \cal A}^{\prime}_N & = - \sum_{j=1}^{N-2} \langle V^{\prime}_{1 \dots j} V^{\prime}_{j+1 \dots N-1} V_N \rangle &\proofNpoint\cr
    & =  - \sum_{j=1}^{N-2} \langle V_{1 \dots j} V_{j+1 \dots N-1} V_N \rangle  = { \cal A}_N.
}$$
Thus the amplitude has the same numerical value for all consistent sets of Berrends-Giele currents. In particular, the worldline prescription gives the same result as the one obtained from the BRST building blocks of \MafraNpoint.

\newsec Discussions

\seclab \secsix

We have developed a prescription for computing tree-level scattering amplitudes 
in the 10D SYM theory using the pure spinor worldline formalism for the superparticle. 
In particular, we have studied the necessary insertions on the worldline, and showed 
that the pinching operators act as contact terms that enhance the BRST properties of 
the kinematical numerators, elucidating how this phenomenon first observed in string 
theory also appears in the particle. 
In particular, one might interpret the multiparticle fields
appearing in the computation of the amplitude as Berends-Giele currents.
Once this is done, the computation of the amplitude boils down to the computation
of these currents, which singles out the known result in terms of the BRST building blocks. 
In this sense, the computation of the amplitude is formulated as a recursive construction
in BRST cohomology, as in \Mafrarecursive. We notice that, in the worldline, this
reformulation arises quite naturally from the integration by parts procedure that we carried over.
The use of the Hamiltonian path integral (as opposed to the usual Lagrangian) in our 
computation of the worldline amplitude was established as a way to avoid ill-defined 
distributions (coming from products of Dirac deltas) in the formalism.

\medskip
Our study provides a foundation for further investigation into the worldline formalism of other supersymmetric theories. A natural next step is the extension to 11D supergravity, using the 11D pure spinor superparticle formalism \refs{\pssupermembrane, \equivalenceguillen, \elevendsimplifiedb, \cederwallequations, \maxmasoncasaliberkovits,\maxthesis, \maxnotesworldline, \tamingbghos, \maxmaorelevend}. We start this research program in our companion papers \refs{\amplitudeeleven, \ghostnumberzero}. Furthermore, the ideas developed in the present manuscript might shed light on the construction of N-point amplitudes for ambitwistor models constructed from their particle counterparts \refs{\nmmax, \maxdiegoone, \maxdiegotwo}. Additionally, exploring loop-level amplitudes within this formalism may offer new insights into the interplay between worldline methods and traditional Feynman diagram techniques. By establishing the effectiveness of the worldline approach for 10D SYM, we pave the way for a broader application of first-quantized techniques in field theory, potentially simplifying amplitude computations in supersymmetric and gravitational theories.

%---------------------------------------------------------------------------------------------------------------------------------------------------------------------------------------------------------------------------------------------------------------------------------------------------------------------------------------------- Appendix ------------------------------------------------------------------------------------------------------------------------------------------------------------------------------------------------------------------------------------------------------------------------------------------------------------------------------------------------------

\bigskip \noindent{\bf Acknowledgements:} MG would like to thank Maor Ben-Shahar, Henrik Johansson, Renann Jusinskas and Oliver Schlotterer for helpful discussions on related topics. MS wants to thanks Nathan Berkovits for helpful discussions. EV wants to thank Kostas Rigatos, Thiago Fleury, Rennan Jusinskas and Humberto Gomez for useful discussions on this topic. We are also grateful to the JHEP referees for their constructive suggestions on the first version of this manuscript. Our thanks also go to ICTP-SAIFR for their hospitality during the workshop on Modern Amplitude Methods for Gauge and Gravity Theories, and to Nordita for organizing Eurostrings 2025, during which this work was finalized. The work of MG was partially funded by the European Research Council under ERC-STG-804286 UNISCAMP, and by the Knut and Alice Wallenberg Foundation under the grant KAW 2018.0162 (Exploring a Web of Gravitational Theories through Gauge-Theory Methods) and the Wallenberg AI, Autonomous Systems and Software Program (WASP). The work of MS was partially supported by  CAPES
(88887.608907/2021-00) and CNPQ (164239/2022-7). The work of EV was supported in part by ICTP-SAIFR FAPESP grant 2019/21281-4 and by FAPESP grant 2022/00940-2.
%OS is indebted to Song He for insightful discussions and collaboration on related topics. 

%%%%%%%%%%%%%%%%%%%%%%%%%%%%%%%%%%%
%%%%%%%%%%%%%%%%%%%%%%%%%%%%%%%%%%%
%%%%%%%%%%%%%%%%%%%%%%%%%%%%%%%%%%%

%------------------------------------------------------------------------------------------------------------------------------------------------------------------------------------------------------------------------------------------------------------------------------------------------------------------------

\medskip

\appendix{A}{Integration by parts}

The last step in the evaluation of the scattering amplitude \Nptfunction\ is the integration over the worldline zero modes. While in the minimal pure spinor formalism this is described ad-hoc by the projection on the measure \measure, the non-minimal pure spinor formalism allows one to justify this result from first principles. One defines the quantum mechanical correlator $\langle {\cal O} \rangle$ as the path integral over the non-minimal variables with insertion ${\cal O}$
\eqnn\correlator
$$\eqalignno{
    { \cal A}_N & = \int {\cal D} \Phi \, {\cal N} e^{- S} {\cal O} , & \correlator
}$$
where ${\cal D} \Phi$ represents the measure for integration over all non-minimal fields and ${\cal N}$ is a regularization factor that ensures that the integral over the bosonic fields is finite. After integrating out $x^m$ and the various non-zero modes using the contractions of equation \wick, one is left with an integral over the zero modes of $\lambda^{\alpha}$, $\bar{\lambda}_{\alpha}$, $r_{\alpha}$ and $\theta^{\alpha}$ 
\eqnn\nonminimalzeromode
$$\eqalignno{
    {\cal A}_N & = \int [d\lambda][d\bar{\lambda}][dr]d^{16}\theta \, {\cal N} f(\lambda, \bar{\lambda}, r, \theta), & \nonminimalzeromode
}$$
for some function $f$. The holomorphic top-form on the respective spaces gives the zero-mode integration measure
\eqnn\measurelambda
\eqnn\measurelambdabar
\eqnn\measurer
$$\eqalignno{
    [d\lambda] \lambda^{\alpha_1} \lambda^{\alpha_2} \lambda^{\alpha_3} & = (\epsilon T^{-1})^{\alpha_1 \alpha_2 \alpha_2}_{\beta_1 \dots \beta_{11}} d \lambda^{\beta_1} \dots d \lambda^{\beta_{11}}, & \measurelambda \cr
    [d\bar{\lambda}] \bar{\lambda}_{\alpha_1} \bar{\lambda}_{\alpha_2} \bar{\lambda}_{\alpha_3} & = (\epsilon T)_{\alpha_1 \alpha_2 \alpha_2}^{\beta_1 \dots \beta_{11}} d \bar{\lambda}_{\beta_1} \dots d \bar{\lambda}_{\beta_{11}}, & \measurelambdabar\cr
    [dr] & = (\epsilon T^{-1})^{\alpha_1 \alpha_2 \alpha_2}_{\beta_1 \dots \beta_{11}} \bar{\lambda}_{\alpha_1} \bar{\lambda}_{\alpha_2} \bar{\lambda}_{\alpha_3} \left( {\partial \over \partial r_{\beta_1}} \right) \dots \left( {\partial \over \partial r_{\beta_{11}}} \right). & \measurer
}$$
where the tensors $(\epsilon T^{-1})^{\alpha_1 \alpha_2 \alpha_2}_{\beta_1 \dots \beta_{11}}$ and $(\epsilon T)_{\alpha_1 \alpha_2 \alpha_2}^{\beta_1 \dots \beta_{11}}$ are symmetric and gamma matrix traceless in $(\alpha_1, \alpha_2, \alpha_3)$ and antisymmetric in $[\beta_1, \dots, \beta_{11}]$. Their explicit form can be found in \NMPS, but this will not be needed here. The presence of the regulator ${\cal N}$ is necessary because the integral over the unbounded bosonic ghosts diverges in its absence. This regulator ${\cal N}$ is BRST trivial and thus cannot change the value of the amplitude. It is given by
\eqnn\Nregulator
$$\eqalignno{
    {\cal N} & = \exp\{ - \{Q, \bar{\lambda}_{\alpha} \theta^{\alpha} \} \} = \exp \{ - \bar{\lambda}_{\alpha} \lambda^{\alpha} - r_{\alpha} \theta^{\alpha}\}, & \Nregulator
}$$
where only the zero-modes of the fields appearing in the last equation contribute. Due to BRST invariance, the amplitude does not depend on the specific choice of regulator. Plugging equation \Nregulator\ into \nonminimalzeromode, the integration furnishes exactly the rule of equation \measure\ \NMPS,\MafraNMoneloop.

The measure \Nregulator\ commutes with the $b$ field defined in \btend. To show this, note first that 
\eqnn\bNregulator
$$\eqalignno{
    [b, \{Q, \bar{\lambda}_{\alpha} \theta^{\alpha}\}] & = - [Q, \{b, \bar{\lambda}_{\alpha} \theta^{\alpha}\}]. & \bNregulator
}$$
since $[H, \bar{\lambda}_{\alpha} \theta^{\alpha}] = 0$. It is simple to check that the (anti)commutator of the physical operators with $\bar{\lambda}_{\alpha} \theta^{\alpha}$ vanishes
\eqnn\physicaloperatorsNregulator
$$\eqalignno{
    [{ \bf \Delta }_m, \bar{\lambda}_{\alpha} \theta^{\alpha}] & = {(\bar{\lambda} \gamma_m r) \over 2 (\bar{\lambda} \lambda)} = 0, \quad \{{\bf A}^m, \bar{\lambda}_{\alpha} \theta^{\alpha}\} = { (\bar{\lambda} \gamma^m \bar{\lambda}) \over 2 (\bar{\lambda} \lambda)} = 0, &\physicaloperatorsNregulator
}$$
where we used the non-minimal pure spinor constraints. This establishes that $[b, {\cal N}] = 0$. It can actually be shown that $[{ \bf A}_m, {\cal N}] = [{ \bf \Delta}^m, {\cal N}] = 0$. This is an important result because it allows one to integrate the action of $b$ by parts in a correlator. This can be demonstrated with ease in the Siegel gauge. A generic kinematical numerator will be given by
\eqnn\integrationbyparts
$$\eqalignno{
    \langle {\cal O}_1 [U^{\prime}_P, {\cal O}_2] \rangle & = \int [d\lambda][d{\bar\lambda}][dr]d^{16}\theta \, {\cal N} \left( {\cal O}_1 [U^{\prime}_P, {\cal O}_2] \right), &\integrationbyparts
}$$
for unspecified operators ${\cal O}_{1,2}$. Writing $U^{\prime}_P = - \hat{{ \bf A}}_m(V^{\prime}_P) \Delta^m + \hat{\Delta}^m(V^{\prime}_P) { \bf A}_m$, we find
\eqnn\integrationbypartstwo
$$\eqalignno{
    \langle {\cal O}_1 [U^{\prime}_P, {\cal O}_2] \rangle & = \langle {\cal O}_1 \left(  - \hat{{ \bf A}}_m(V^{\prime}_P) \hat{\Delta}^m({\cal O}_2) + \hat{\Delta}^m(V^{\prime}_P) { \bf A}_m({\cal O}_2)\right) \rangle. \cr
     & = \langle \left( \hat{{ \bf A}}_m(V^{\prime}_P) \hat{\Delta}^m({\cal O}_1) - \hat{\Delta}^m(V^{\prime}_P) { \bf A}_m({\cal O}_1)\right) {\cal O}_2 \rangle. \cr
     & = - \langle [U^{\prime}_P, {\cal O}_1] {\cal O}_2 \rangle, & \integrationbypartstwo
}$$
where we integrated by parts in the second line, and used the fact that ${ \bf A}_m$ is fermionic while ${ \bf \Delta}^m$ is bosonic. We have also used the Siegel gauge condition $b_0(V^{\prime}_P) = 0$, and the fact that the physical operators commute with the measure ${\cal N}$. Finally, we have also used the fact that $N^{mn}$ acts as a total derivative on pure spinor space,
\eqnn\Nmntotalderivative
$$\eqalignno{
    \int [d\lambda] \; \lambda^{\alpha} (\gamma_{mn})_{\alpha}^{\;\beta} \partial_{\lambda^{\beta}} [f(\lambda)] & = 0, & \Nmntotalderivative
}$$
provided that the function $f$ is sufficiently well-behaved at the boundaries. 

In Siegel gauge, this implies that the kinematical numerators explicitly satisfy generalized Jacobi identities, as show in \maxmaor. Indeed, the kinematical numerators appearing in the tree amplitude in Siegel gauge are generally given by
\eqnn\numeratorSiegelgauge
$$\eqalignno{
    & \langle b_0(V^{\prime}_A V^{\prime}_B) V^{\prime}_C V^{\prime}_D \rangle. & \numeratorSiegelgauge
}$$
Using the fact that $b_0$ is a quadratic differential operator, we find
\eqnn\numeratorSiegelgaugetwo
$$\eqalignno{
    b_0(V^{\prime}_A V^{\prime}_B V^{\prime}_C) & = b_0( V^{\prime}_A V^{\prime}_B) V^{\prime}_C + b_0( V^{\prime}_B V^{\prime}_C) V^{\prime}_A + b_0( V^{\prime}_C V^{\prime}_A) V^{\prime}_B. & \numeratorSiegelgaugetwo
}$$
Integrating by parts, we obtain the following generalized Jacobi identities
\eqnn\jacobiidentities
$$\eqalignno{
    \langle b_0(V^{\prime}_A V^{\prime}_B) V^{\prime}_C V^{\prime}_D  \rangle  + \langle b_0(V^{\prime}_B V^{\prime}_C) V^{\prime}_A V^{\prime}_D \rangle + \langle b_0(V^{\prime}_C V^{\prime}_A) V^{\prime}_B V^{\prime}_D \rangle & = 0. & \jacobiidentities
}$$

\seclab\appendixb

\appendix{B}{BRST building blocks}

In \MafraNpoint, the contractions of the N-point function are organized using the BRST
building blocks. They behave like the composite operators $\hat U_{i_1}\cdots \hat U_{i_k}(V)$ in the sense that they describe multiparticle states and obey the BRST cohomology of equation \allrankV. But they also satisfy symmetries that are reminiscent of the generalized Jacobi identities and can be used to construct the Berends–Giele currents of SYM. In this appendix, we will briefly explain their construction. The interested reader should consult the original reference \MafraNpoint.

\medskip
\noindent The composite fields that appear in the worldline prescription \Nptfunction\ are given by strings of integrated vertices $U_{P_i}$ acting on the unintegrated vertices $V_i, \hat U_{P_1}\cdots\hat U_{P_k}(V_i)$. At the first level, we obtain the two-particle composite field
\eqnn\UV
$$\eqalignno{
&\hat{U}_i(V_j) = - \Big( (A_i \cdot k_j) V_j + (W_i \gamma^m \lambda) A_{jm} + Q( W_i A_j) \Big). &\UV
}$$

The worldline prescription also forces us to consider the composite fields representing a larger number of particles, such as
\eqnn\UUV
$$\eqalignno{
&\hat U_3 \hat U_2 V_1,\cr
&\hat U_{[23]}V_1,&\UUV\cr 
&\hat U_4\hat U_3 \hat U_2 V_1,\cr
&\ \ \ \ \ \vdots
}$$

The string theory computation contains equivalent objects. These are the composite fields $L_{ijkn...}$, defined as the single pole contraction of integrated and unintegrated vertices. At first order, we define the composite field $L_{21}$ is computed by taking the single pole in the OPE of $U_2$ colliding with $V_1$,
\eqnn\Lonetwo
$$\eqalignno{
& L_{21}(z_1) = \lim_{z_2\to z_1}(z_2-z_1)U_2(z_2)V_1(z_1).&\Lonetwo
}$$
At higher orders, the composite field $L_{ijkn...}$ is defined by taking $z_i\to z_j$, $z_k\to z_n$, etc in the OPE. These can be obtained recursively. For instance,
\eqnn\Ln
$$\eqalignno{
& L_{2131\cdots n1} = \lim_{z_n\to z_1}(z_n-z_1)U_n(z_n)L_{2131\cdots}(z_1).&\Ln
}$$
The equivalence between the composite vertices of the particle and the $L_{ijkn...}$ can be established with ease since the extraction of the single pole in $U(z){\cal O}(0)$ where ${\cal O}$ has conformal weight zero is equivalent to the particle contraction $[U, {\cal O}]$. The
composite fields satisfy the BRST cohomology
\eqnn\QLoneap
\eqnn\QLtwoap
$$\eqalignno{
&QL_{21} = s_{12}V_1V_2,&\QLoneap\cr
&QL_{2131} = s_{123}L_{21}V_3 + s_{12}\Big[ L_{31}V_2 + V_1L_{32} - L_{21}V_3 \Big],&\QLtwoap
}$$
and similarly for higher-order vertices. This can be obtained by using the descent equation $[Q, U] = \partial V$. This BRST cohomology allows one to assign a diagrammatic interpretation to the composite fields, as done in \MafraNpoint. This form of the BRST cohomology is fixed by the requirement of BRST invariance of the tree-amplitude. Let us see how this works out for the color-ordered 4-pt function. If we write generally
\eqnn\colorfourpoint
$$\eqalignno{
& {\cal A}_4 = {n_s\over s} + {n_t\over t}, &\colorfourpoint
}$$
then this can only satisfy $Q{\cal A}_4 = 0$ for arbitrary values of the external momenta if $Qn_s = s{\cal O}$ and $Qn_t = −t{\cal O}$ for some ${\cal O}$. This is exactly the general structure of BRST cohomology of the composite fields given by equation \QLoneap, so it is intuitive that the composite fields correspond to the kinematical numerators of the tree-amplitude.

Different kinematical numerators have different properties. It is of interest to enhance the properties of the kinematical numerators by performing generalized gauge transformations, which is equivalent to the cancellation of contact terms between diagrams. The contact terms correspond to the BRST exact pieces of the composite fields. 

The BRST exact piece of $L_{21}$ can be found by direct inspection to be the symmetric part,
\eqnn\contactL
$$\eqalignno{
& L_{12} + L_{21} = -QD_{12}. &\contactL
}$$
where $D_{12} = (A_1 \cdot A_2 + W_1A_2 + W_2A_1)$. In the string, this might be established with ease by considering the BRST charge acting on the OPE of two integrated vertices $U_1(z)U_2(0)$. By first taking the double pole and then acting with $Q$, we obtain $−[Q, D_{12}](0)/z$ where $−D_{12}$ is the (symmetric) double pole of the OPE. By first acting with the BRST charge and then taking the double pole, we obtain $(L_{21} + L_{12})(0)/z^2$, and since the order of operations cannot alter the result, this establishes that $L_{\{12\}}$ is BRST exact on the string. For the superparticle, one notes that the consistency condition \ranktwoD\ implies that $\hat U_1(V_2) + \hat U_2(V_1)$ must be BRST trivial.

When we subtract the BRST exact pieces, we define the BRST building blocks $T_{ij\cdots}$. For instance,
\eqnn\Tijap
$$\eqalignno{
& T_{12} = L_{[12]} = L_{21} - L_{\{21\}} = L_{21} + \half QD_{12}. &\Tijap
}$$
This is the rank-2 BRST building block. It satisfies the same BRST cohomology as $L_{21}$, but has the advantage of also being anti-symmetric,
\eqnn\Tantisymap
$$\eqalignno{
& T_{\{12\}} = 0. &\Tantisymap
}$$
The higher-order BRST building blocks are constructed similarly. From the BRST cohomology of the rank-2 composite field, we obtain
\eqnn\QLijklap
$$\eqalignno{
QL_{2131} = & s_{123}T_{12}V_3 + s_{12}\Big[ T_{13}V_2 + V_1T_{23} - T_{12}V_3 \Big] \cr
& -\half Q\left( s_{123}D_{12}V_3 + s_{12}\Big[ D_{13}V_2 - V_1D_{23} - D_{12}V_3 \Big]\right).&\QLijklap
}$$
This implies that
\eqnn\preTijklap
$$\eqalignno{
&  Q\left( L_{2131} + \half s_{123}D_{12}V_3 + \half s_{12}\Big[ D_{13}V_2 - V_1D_{23} - D_{12}V_3 \Big]\right) \cr
& =  s_{123}T_{12}V_3 + s_{12}\Big[ T_{13}V_2 + V_1T_{23} - T_{12}V_3 \Big].&\preTijklap
}$$
We define $T_{123}$ as the object inside parenthesis, but stripped of its BRST exact pieces,
\eqnn\Tijklap
$$\eqalignno{
& T_{123} =  L_{2131} + \half s_{123}D_{12}V_3 + \half s_{12}\Big[ D_{13}V_2 - V_1D_{23} - D_{12}V_3 \Big] + Q(\cdots). &\Tijklap
}$$
The construction of the BRST building blocks goes on in this iterative manner. The BRST building blocks satisfy the BRST cohomology appropriate for kinematical numerators, e.g.
\eqnn\QTijap
\eqnn\QTijkap
$$\eqalignno{
& QT_{12} = s_{12}V_1V_2, &\QTijap\cr
& QT_{123} = s_{123}T_{12}V_3 + s_{12}\Big[ T_{13}V_2 + V_1T_{23} - T_{12}V_3 \Big], &\QTijkap
}$$
and similarly for higher rank BRST building blocks. The subtraction of the BRST exact pieces also enhances the symmetry properties of these objects, e.g.
\eqnn\fap
\eqnn\fapp
$$\eqalignno{
& T_{\{12\}} = 0, &\fap\cr
& T_{\{12\}3} = T_{\{123\}} = 0.  &\fapp
}$$
The composite fields obtained in the worldline formalism are equivalent to the $L_{ijkn\cdots}$ composite fields, as argued. In particular, they satisfy the same BRST cohomology as the $L_{ijkn\cdots}$ and the BRST building blocks $T_{ijk\cdots}$. By arguments of uniqueness in the BRST cohomology, we might relate any such set of operators by the subtraction of contact terms and BRST exact pieces. As we have seen, the worldline amplitude is naturally written in terms of the $\hat U_i \cdots \hat U_j V_k$ composite fields. The pinching operators conspire to kill off the BRST exact pieces of these fields, enhancing the symmetry properties of the numerators. Crucially, this procedure allows one to bootstrap the tree-level amplitude from BRST invariance once one computes the composite fields $\hat U_i \cdots \hat U_j V_k$, as we did.

\listrefs

\bye